\documentclass[twocolumn,amssymb,eqsecnum,aps,pra]{revtex4-2}
\par
\usepackage{amsmath}
\usepackage{enumerate}
\usepackage{graphicx}
\begin{document}
\def \bbeta {{\beta}\hskip -5.6pt {\beta}}
\def \brho {{\rho} \hskip -4.5pt { \rho}}
\def \bnab {{\bf\nabla}\hskip-8.8pt{\bf\nabla}\hskip-9.1pt{\bf\nabla}}
\title{Alternative formulation of the macroscopic field equations in a linear magneto-dielectric medium: Lagrangian field theory and spacetime setting}
\author{Michael E. Crenshaw}
%\email{michael.crenshaw4.civ@mail.mil
\affiliation{US Army Combat Capabilities Development Command (DEVCOM) - Aviation and Missile Center, Redstone Arsenal, AL 35898, USA}
\begin{abstract}
\par
A transparent linear magneto-dielectric material in free space that is
illuminated by a finite quasimonochromatic field is a thermodynamically
closed system, definitively, regardless of what field and material
subsystems that one defines.
The energy--momentum tensor that is formally derived from the
Maxwell--Minkowski field equations is inconsistent with both angular
and linear momentum conservation in this closed system; this very
solid fact is the foundational and continuing issue of the
Abraham--Minkowski controversy.
The extant resolution of the Abraham--Minkowski dilemma is to treat
Maxwellian continuum electrodynamics as being a subsystem and to
write the total energy--momentum tensor as the sum of a Maxwellian
electromagnetic subsystem energy--momentum tensor and a
phenomenological material subsystem energy--momentum tensor.
We prove that fundamental principles of physics are violated by
Maxwellian continuum electrodynamics and that fundamental principles
of physics are violated by Maxwellian continuum electrodynamics
supplemented by the material subsystem conjecture.
We use field theory to derive legitimate equations for macroscopic
electromagnetic fields in a transparent linear magneto-dielectric
medium.
The new field equations are a part of a self-consistent formulation
of macroscopic electrodynamics, conservation laws, special
relativity, and invariance in a continuous linear medium.
In the new formulation, the temporal and spatial coordinates are
renormalized by the continuous linear medium instead of the
permittivity and permeability being carried as independent
material parameters. Then an isotropic, homogeneous, flat,
four-dimensional, continuous, linear, non-Minkowski spacetime
is the proper setting for the continuum electrodynamics of a
simple linear medium in which the effective speed of light is
$c/n$ and each medium will be associated with a different
spacetime.
%\hfill
\end{abstract}
\date{\today}
\maketitle
\par
\section{Introduction}
\par
%The microscopic Maxwell equations for electromagnetic fields in the
%vacuum of free space are fundamental laws of physics.
%The macroscopic Maxwell equations for electromagnetic fields in a
%linear magneto-ielectric medium are not fundamental; they
%are phenomenological.
%\par
Continuum electrodynamics can be defined as a formal theory 
whose axioms are the Maxwell--Minkowski equations (macroscopic Maxwell
equations), the constitutive relations, and the definitions of the 
fields in terms of the vector potential.
Theorems of continuum electrodynamics are formally generated by the
operations of algebra and calculus on the axioms.
In particular, the electromagnetic conservation law 
(see Eq.~(\ref{EQq2.15})) is a theorem of the Maxwell--Minkowski
equations and constitutive relations in the limit that the gradient
Minkowski four-force density is negligible.
The formal derivation of the electromagnetic conservation law
also derives the Minkowski energy--momentum tensor and the Minkowski
momentum density.
\par
We address a contradiction between the linear momentum
conservation properties of two theorems of formal continuum
electrodynamics; the electromagnetic conservation law and the wave
equation.
Specifically, the vanishing four-divergence of the Minkowski 
energy--momentum tensor for a quasimonochromatic field normally
incident on a gradient-index antireflection coated transparent linear
magneto-dielectric medium proves, via the (local) electromagnetic
conservation law, that the Minkowski linear momentum is
conserved \cite{BIPfei,BIBrevCons,BIBrevVec,BIBrevik,BIWang4Vec},
while global conservation analyses of the same system using the wave
equation prove that the Minkowski linear momentum in the medium
is greater than the vacuum-incident momentum by a factor of the
refractive index $n$ \cite{BIPfei,BIGord,BIFofn1,BIFofn2}.
\par
The formal theory is a precise axiomatic system and we must
acknowledge that the disproof of a theorem of the formal theory,
via contradiction between the electromagnetic conservation law and
the wave equation, proves that the axioms, the macroscopic
Maxwell--Minkowski equations, constitutive relations, and
vector potential relations, are false.
Instead, scientists assume that the model system is incomplete,
ostensibly requiring the inclusion of a material-motion subsystem and a
heuristic way to couple the phenomenological material subsystem to the
electrodynamic subsystem in order to reconcile Maxwellian electrodynamic
theory with the violated physical principles 
\cite{BIRL,BIBoydMil,BIPfei,BIAMC4,BIKemplatest,BIPenHaus,BIGord,BIBarn,BIBarnLou,BIKranys,BIObukPLA,BIBrevCons}.
\par
The postulated incompleteness of the Minkowski energy--momentum
tensor proves the incompleteness of the electromagnetic conservation
law and thereby proves the incompleteness of the axioms from which the 
electromagnetic conservation law is derived as a theorem.
Given that the electromagnetic conservation law, a formal theorem
of the Maxwell-Minkowski field equations, results in
long-known, scientifically reported
\cite{BIRL,BIBoydMil,BIPfei,BIAMC4,BIKemplatest,BIPenHaus,BIGord,BIBarn,BIBarnLou,BIKranys,BIObukPLA,BIBrevCons},
and easily verified errors in the momentum, the question as to
why the incomplete macroscopic Maxwell--Minkowski field equations are
regarded as laws of physics, still, and continue to be treated as
fundamental equations in textbooks and other scientific publications
is a disturbing issue of scientific philosophy.
\par
The foundational issue \cite{BIMin,BIAbr} of the century-old
Abraham--Minkowski controversy
\cite{BIRL,BIBoydMil,BIPfei,BIAMC4,BIKemplatest,BIPenHaus,BIGord,BIBarn,BIBarnLou,BIKranys,BIObukPLA,BIBrevCons}
is that the non-symmetric Minkowski energy--momentum tensor
is not consistent with conservation of angular momentum.
Originally an issue of angular momentum
conservation \cite{BIMin,BIAbr}, nowadays the Abraham--Minkowski
controversy is typically characterized as a question about the
physical meaning and usage of two different electromagnetic linear 
momentum formulas, the Minkowski electromagnetic momentum and the
Abraham electromagnetic momentum, and the corresponding material
subsystem momentums
\cite{BIRL,BIBoydMil,BIPfei,BIAMC4,BIKemplatest,BIPenHaus,BIGord,BIBarn,BIBarnLou,BIKranys,BIObukPLA,BIBrevCons,BIBrev}.
\par
The `modern' resolution of the Abraham--Minkowski controversy,
reviewed in Sec. II, is to
posit the existence of a material subsystem with a material subsystem 
energy--momentum tensor (or a material subsystem momentum)
\cite{BIRL,BIBoydMil,BIPfei,BIAMC4,BIKemplatest,BIPenHaus,BIGord,BIBarn,BIBarnLou,BIKranys,BIObukPLA,BIBrevCons}.
The coupling between the electromagnetic and material subsystems is
derived by global conservation of energy, global conservation of
linear momentum, and conservation of angular momentum.
Then we can write a total (field plus material) energy--momentum tensor
(or the total momentum) \cite{BIPfei,BIJMP,BIGord,BIBarn,BIBarnLou}.
Except, we prove that using the total (field plus material) 
energy--momentum tensor \cite{BIPfei,BIJMP} in the electromagnetic
conservation law repairs global linear momentum
conservation and angular momentum conservation, by hand, but
violates other physical principles, i.e. relativity and the Poynting
theorem, that are well-known to be satisfied by the Maxwell--Minkowski
field equations.
Then the conservation law is false without the material subsystem
conjecture and it is false with the material subsystem conjecture.
\par
Starting with Abraham \cite{BIAbr} in 1909, manifold disproofs
\cite{BIRL,BIBoydMil,BIPfei,BIAMC4,BIKemplatest,BIPenHaus,BIGord,BIBarn,BIBarnLou,BIKranys,BIObukPLA,BIBrevCons}
of the electromagnetic conservation law
also \textit{disprove the Maxwell--Minkowski equations,} the axioms
from which the conservation law is formally constructed as a theorem
that becomes an identity in the limit that a transient is neglected.
Then the falsification of the Maxwell-Minkowski equations is a 
century-old matter of abstract algebra.
Because the literature of the Abraham--Minkowski controversy contains
a great amount of deflection about this point, we disprove several
hypothetical resolutions of the issue in Sec.~II, even though the
disproof of the local electromagnetic conservation law 
is sufficient.
\par
Minkowski spacetime is a model for the vacuum \cite{BIGiu2}.
The spacetime conservation laws, Sec. III, are fundamental physical
principles of Minkowski spacetime for the propagation of the unimpeded
(no external forces), inviscid, incoherent, incompressible flow of
non-interacting particles (nonpolar molecules, dust particles, or
photons) in the continuum limit (fluid or light field (light fluid))
through an otherwise empty vacuum \cite{BIFox}.
The conservation laws of Minkowski spacetime are fixed and
immutable for any countable system obeying the specified conditions,
for example, fluid flow in the vacuum or light flow in the vacuum.
However, those laws change with physical circumstances.
For example, the Navier--Stokes equation is a version of the
spacetime conservation law that has been modified for a viscous
flow through the otherwise empty vacuum.
\par
Although the vacuum can be modeled as a Minkowski spacetime, 
we show that the effective medium description of a transparent
simple linear magneto-dielectric medium in the continuum limit
corresponds to a linear, isotropic, homogeneous, flat,
four-dimensional, continuous non-Minkowski `material' spacetime
in which the effective speed of light is $c/n$.
Because the unimpeded, inviscid, incoherent flow of a spatially
compressed (compressed and de-compressed at the boundaries of the
material, but not otherwise compressible) light field is
traveling through a region of space that is not a vacuum, we must
develop the physical laws that apply in the `material' spacetime.
The Maxwell--Minkowski equations are readily derived by using 
Lagrangian field theory in Minkowski spacetime \cite{BIGold,BICohen}.
In Secs. IV and V, we develop field theory for a non-empty region
of space that responds linearly to electromagnetic radiation with
a speed of light $c/n$.
\par
The current author performed a simple application of special relativity
using inertial reference frames moving uniformly along the surface
of a large transparent linear dielectric in Ref.~\cite{BIAJP}.
It was shown that Einstein's special relativity manifests differently
for observers on opposite sides of the interface:
\textit{i}) An observer on the vacuum side of the interface uses
boundary conditions to describe events that occur inside the
dielectric using Laue's \cite{BILaue} dielectric special relativity
with a vacuum Lorentz factor and a velocity-dependent index of
refraction.
\textit{ii}) An observer on the dielectric side of the interface
finds that the application of Einstein's relativity in a dielectric
is best described by Rosen's \cite{BIRosen} dielectric special
relativity in terms of a non-Lorentz (but Lorentz-like) factor
that contains the permittivity of the dielectric and does not
depend on the velocity of the dielectric.
In Sec. VI, the derivation of Ref.~\cite{BIAJP} is applied to a 
simple linear magneto-dielectric medium and the result demonstrates 
that the Lorentz-like factor for the observer inside the continuous
medium depends on the permittivity and permeability through the index
of refraction.
An observer inside an arbitrarily large continuous linear, isotropic,
homogeneous, magneto-dielectric medium determines that the velocity of
light $c/n$ is independent of the velocity of
the source in accordance with the Principle of Relativity.
\par
There are additional optical processes that need to be re-evaluated
when they occur in a transparent, isotropic, homogeneous,
continuous linear medium, instead of in the vacuum.
In Sec. VII, we prove a linear, isotropic, homogeneous, flat,
four-dimensional, continuous `material' spacetime in which the
temporal coordinate is normalized by the inverse of the square
root of the permittivity and the spatial coordinates are normalized
by the square root of the permeability.
As has been known \cite{BIFinn}, Lorentz invariance is not of
symmetry of a linear medium.
In Sec. VIII, we establish a refractive-index-dependent invariance for the
medium-dependent non-Minkowski spacetime of a continuous linear medium.
This result implies that Laue's theorem \cite{BIGiu,BIWang2}
and Noether's theorem \cite{BINoether} are re-defined
for a simple linear medium by the new invariance principle, however,
we focus on the more common electrodynamic principles and we do
not derive these two theorems here.
In Sec. IX, we construct a tensor formulation of continuum
electrodynamics as theorems of the new field equations and note
that the energy--momentum tensor is diagonally symmetric and that the
electromagnetic energy and electromagnetic momentum are locally
and globally conserved without the need for a material-motion
subsystem.
Finally, the new theory of continuum electrodynamics is shown to be
consistent with the Balazs \cite{BIBalazs} thought experiment and the 
Jones--Richard \cite{BIExp} mirror experiment.
It is shown that the uniform velocity of the center of mass-energy
theorem depends on a constant mass-energy density.
When applied to light propagation, the uniform velocity of the center
of mass-energy theorem \cite{BIExp} must be modified to account for
the change in the volume occupied by the energy and momentum of the
field that is accompanied by the corresponding change in the energy
and momentum density, Sec. X.
\par
\section{Continuum Electrodynamics}
\par
The model system consists of a finite quasimonochromatic 
electromagnetic field and a block of
simple linear magneto-dielectric material located in a
large finite volume ${\Sigma}$ of free space. 
We define a simple linear medium as a transparent, isotropic,
homogeneous, continuous linear medium that has no resonances
near the center frequency of the quasimonochromatic field;
the material is ``effectively dispersionless at frequencies of
interest'' \cite{BIBoydMil} (There is some ambiguity in the
terminology used in the literature because dispersion is actually
treated in lowest-order in the `dispersionless' cases.
The refractive index $n$ depends on the frequency of the field,
but the frequency is treated as being constant for the duration of
the quasimonochromatic field.)
The material is initially at rest in the local frame.
Unless the radiation is of extraordinary intensity and duration,
the velocity of the material in the local frame will be
non-relativistic and neglecting the effects of the material motion
on the refractive index is an ``extremely accurate approximation
indeed'' \cite{BIObukPLA}.
Then dispersion and velocity-dependence can be treated in lowest order
such that the permittivity, permeability, and the refractive index
can be represented by real constants.
The values of these constants depend on the properties of the material
and depend on the center frequency of the exciting quasimonochromatic
field (dispersion is treated in lowest-order).
The model system is the principal model of a simple linear medium
that is extensively used in continuum electrodynamics, for example,
the derivation of the Fresnel
relations \cite{BIJackson,BIGriff,BIZangwill}.
The stationary `dispersionless' limit is implicitly and explicitly
used in most lowest-order expositions of the Abraham--Minkowski
controversy \cite{BIBoydMil}.
\par
The electromagnetic theory is developed using vector and tensor
formulations.
For purposes of illustration, to compare magnitudes, for example,
propagation of the field is discussed using the plane-wave limit.
The plane-wave limit is a common abstraction with well-known
characteristics that allows paraxial problems to be treated
in lowest-order with one spatial dimension, not to be confused
with the assumption of uniform plane waves that are nonphysical
due to their infinite energy.
The plane-wave limit is used in the typical derivation of the
Fresnel relations and many other elementary problems of continuum
electrodynamics \cite{BIJackson,BIGriff,BIZangwill}.
The plane-wave limit is explicitly and implicitly used in many
lowest-order expositions of the Abraham--Minkowski controversy.
\par
The model quasimonochromatic field is initially in the vacuum and
has a constant amplitude except for a short smooth turn-on transition
and a short smooth turn-off transition.
The field propagates toward and then enters the transparent,
isotropic, homogeneous linear medium at normal incidence
through a gradient-index antireflection coating.
The field re-enters the vacuum through the gradient-index
antireflection coating on the opposite side of the medium.
The system, as defined, is obviously closed.
In particular, any reflected field and whatever material motion
that is imparted by the interaction with the field are part of
the closed system along with the refracted and transmitted fields.
Conservation laws can be applied to the thermodynamically closed
system \cite{BIBrevCons}.
\par
There is no scientific error in deriving theoretical results for
a limiting case in a closed system.
Once the theoretical results are derived for the limiting case 
(quasimonochromatic field, lowest-order dispersion, stationary medium,
no sources or sinks, etc), the theory can be extended to more
detailed models.
\par
Brevik \cite{BIBrevCons,BIBrevik} and Wang \cite{BIWang4Vec} use the
vanishing four-divergence of the Minkowski energy--momentum tensor
(see Eqs.~(\ref{EQq2.14a}) and (\ref{EQq2.15})) as a local
conservation law \cite{BIGiu} to prove that the Minkowski energy
\begin{equation}
U_M=\int_{\Sigma}\frac{{\bf D}\cdot{\bf E}+{\bf B}\cdot{\bf H}}{2} dv
\label{EQq2.01}
\end{equation}
and the Minkowski linear momentum
\begin{equation}
{\bf G}_M=\int_{\Sigma}\frac{{\bf D}\times{\bf B}}{c}dv
\label{EQq2.02}
\end{equation}
form a Lorentz four-vector $(U_M,{\bf G}_M)$ in the limit that the
Minkowski four-force density ${\bf f}^{\alpha}_M$ that is associated
with the gradient-index antireflection coating can be neglected.
This is considered to be a resolution of the Abraham--Minkowski
controversy because the elements of a Lorentz four-vector are globally
conserved \cite{BIWang4Vec,BIGiu}.
\par
Except, that is not the case here.
Although the Abraham--Minkowski dilemma was originally about
conservation of angular momentum, it was well-known, almost from the
outset of the controversy, that the Minkowski linear
momentum ${\bf G}_M$ is not globally conserved 
\cite{BIPfei,BIFofn1,BIFofn2,BIGord}
thereby contradicting the determination 
\cite{BIBrevCons,BIBrevik,BIWang4Vec}
that $(U_M,{\bf G}_M)$ is a Lorentz four-vector.
\par
That being said, it is common practice to dismiss the global
conservation problem with the linear momentum by deeming the
violation of global momentum conservation to be negligible based
on the vanishing four-force density ${\bf f}^{\alpha}_M$ that
appears as the right-hand side of the electromagnetic conservation
law (see Eq.~(\ref{EQq2.14a})).
Although the practice is scientifically countenanced by appealing to
the electromagnetic conservation law (see Eq.~(\ref{EQq2.15})), the
deduction contradicts the long--known, scientifically
reported \cite{BIPfei,BIFofn1,BIFofn2,BIGord} and easily verified
fact that the Minkowski momentum in the material is the 
momentum of the field that is incident from the vacuum 
multiplied by a non-negligible factor of $n$.
\par
\textit{A substantive contradiction exists between the
(local) electromagnetic conservation law and the global
conservation law.}
Adopting either the (local) electromagnetic conservation law or
global conservation dictates the direction of the analysis and
disproves the other.
Although both aspects of the contradiction appear in the scientific
literature, they typically appear separately thereby avoiding obvious
contradictions.
%%%%%Plus, you can cite references that support whichever side of
%%%%%the argument that you prefer.
In their detailed, comprehensive review article,
Pfeifer, Nieminen, Heckenberg, and Rubinsztein-Dunlop \cite{BIPfei},
present both sides of the issue but, due to the structure of a
review article, the two results appear in different sections of
the article with the global result of a factor of $n$ difference
in the linear momentum being mentioned in Sec.~III while the
Minkowski momentum is described as (almost) conserved in Sec.~VI-A
in connection with the electromagnetic conservation law.
\par
Next, we review the details of the argument using the familiar
Maxwell--Minkowski formulation of macroscopic electrodynamics.
Continuum electrodynamics can be described as a formal theory
whose axioms are the Maxwell--Minkowski equations,
\begin{subequations}
\begin{equation}
\nabla\times{\bf H}- \frac{\partial {\bf D}}{\partial (ct)}=
\frac{{\bf J}_f}{c}
\label{EQq2.03a}
\end{equation}
\begin{equation}
\nabla\times{\bf E}+ \frac{\partial {\bf B}}{\partial (ct)}=0
\label{EQq2.03b}
\end{equation}
\begin{equation}
\nabla\cdot{\bf D}=\rho_f
\label{EQq2.03c}
\end{equation}
\begin{equation}
\nabla\cdot{\bf B}=0\; ,
\label{EQq2.03d}
\end{equation}
\label{EQq2.03}
\end{subequations}
and constitutive relations,
\begin{subequations}
\begin{equation}
{\bf D}=\varepsilon {\bf E}
\label{EQq2.04a}
\end{equation}
\begin{equation}
{\bf B}=\mu {\bf H}
\label{EQq2.04b}
\end{equation}
\begin{equation}
n=\sqrt{\varepsilon\mu} \;,
\label{EQq2.04c}
\end{equation}
\label{EQq2.04}
\end{subequations}
for the macroscopic fields ${\bf E}$, ${\bf B}$, ${\bf D}$, and
${\bf H}$ in a simple linear magneto-dielectric medium.
Later, the use of the wave equation will cause us to treat the
relations between the vector potential ${\bf A}$ and the macroscopic
fields (see Eqs.~(\ref{EQq2.17})) as axioms, as well.
\par
The free charge density $\rho_f$ and the free current density
${\bf J}_f$ are macroscopic parameters.
Also, $\varepsilon$ is a continuum abstraction of
the electric permittivity, $\mu$ is a continuum abstraction of the
magnetic permeability, and $n$ is the macroscopic refractive index.
The physical system, as we have defined it, allows us to treat the
material parameters $\varepsilon$, $\mu$, and $n$ in lowest order
as depending on the center frequency of the quasimonochromatic
field but are otherwise single-valued real constants
\cite{BIJackson,BIGriff,BIZangwill}.
\par
Describing the theoretical viewpoint of physics,
Rindler \cite{BIRindler} states ``a physical theory is an abstract
mathematical model (much like Euclidian geometry) whose applications
to the real world consist of correspondences between a subset of it
and a subset of the real world''.
\textit{Continuum electrodynamics is constructed as a formal theory
in this abstract mathematical framework} by performing operations of
algebra and calculus on the axioms.
If any theorem of Eqs.~(\ref{EQq2.03}), (\ref{EQq2.04}), and 
(\ref{EQq2.17}) is proven false, then one or more of the axioms
are proven false and all other theorems that are derived from the
axioms are unproven.
\par
Experimentalists \cite{BIPfei,BIBrev} have a different viewpoint
and are concerned about including the full set of physical effects
that might affect measurements because their experimental conditions
are not usually as pristine as a theoretical model.
Real-world effects, like damping, material motion, higher orders of
dispersion, electrostriction, non-linearity, etc, can be important
in a general setting, but these effects are obviously not going to
fix the essential contradiction between the (local) electromagnetic 
conservation law and global conservation in the physical theory of
the specified model system.
Adding these higher-order effects to a provably flawed
model is a meritless appeal to complexity in the face of a 
contradiction between theorems of the Maxwell--Minkowski equations.
Those higher-order effects can be incorporated later to align
the theory with experiments over a broad range of conditions once the
contradiction is resolved.
\par
Derivations of the electromagnetic momentum density continuity equation
(momentum conservation law) typically begin with the Lorentz force law
\cite{BIBoydMil,BIJackson,BIGriff,BIZangwill}.
The free charge momentum density ${\bf p}_{mech}$ imparted by the field
to a distribution of free charges in the continuum limit can be
calculated by postulating the Lorentz force density 
\cite{BIBoydMil,BIJackson,BIGriff,BIZangwill}
\begin{equation}
\frac{d{\bf p}_{mech}}{dt}={\bf f}_L=
\rho_f{\bf E}+ \frac{{\bf J}_f}{c}\times{\bf B} 
\label{EQq2.05}
\end{equation}
as a physical law \cite{BIManx,BIMansurx}.
The sources are eliminated in favor of the fields using the Gauss law,
Eq.~(\ref{EQq2.03c}), to eliminate $\rho_f$ and using the
Maxwell--Amp\`ere law, Eq.~(\ref{EQq2.03a}), to eliminate ${\bf J}_f$.
Then the momentum density ${\bf p}_{mech}$ imparted to the
free-charge density can be calculated by
integrating \cite{BIBoydMil,BIJackson,BIGriff,BIZangwill}
\begin{equation}
\rho_f{\bf E}+ \frac{{\bf J}_f}{c}\times{\bf B} =
(\nabla\cdot {\bf D}){\bf E}+ \left (\nabla\times{\bf H}
-\frac{1}{c}\frac{\partial{\bf D}}{\partial t}\right )
\times{\bf B} \, .
\label{EQq2.06}
\end{equation}
Substituting the calculus identity
\begin{equation}
\frac{\partial}{\partial t}({\bf D}\times{\bf B}) =
\frac{\partial{\bf D}}{\partial t}\times{\bf B} +
{\bf D}\times \frac{\partial{\bf B}}{\partial t} \, ,
\label{EQq2.07}
\end{equation}
Faraday's law, Eq.~(\ref{EQq2.03b}), Thompson's law,
Eq.~(\ref{EQq2.03d}), and Gauss's law into Eq.~(\ref{EQq2.06})
yields the momentum continuity
equation \cite{BIJackson,BIGriff,BIZangwill}
$$
\rho_f{\bf E}+\frac{{\bf J}_f}{c}\times{\bf B} =
(\nabla\cdot{\bf D}){\bf E} +
$$
\begin{equation}
(\nabla\cdot{\bf B}){\bf H} 
-{\bf D}\times(\nabla\times {\bf E})
-{\bf B}\times(\nabla\times{\bf H})
-\frac{1}{c}\frac{\partial}{\partial t}({\bf D}\times{\bf B}) \, .
\label{EQq2.08}
\end{equation}
%%Eq.~(\ref{EQq2.08}) is also known as the momentum conservation law.
The textbook derivation is simple and the steps have obvious physical
meaning.
The textbook derivation is not as rigorous as we would like because
we are unnecessarily postulating the Lorentz force density law,
Eq.~(\ref{EQq2.05}) 
\cite{BIBoydMil,BIJackson,BIGriff,BIZangwill,BIManx,BIMansurx}.
\par
We propose an alternative derivation of the energy and momentum
continuity equations as formal theorems of the Maxwell--Minkowski
equations.
We take the scalar product of Eq.~(\ref{EQq2.03b}) with ${\bf H}$ and
the scalar product of Eq.~(\ref{EQq2.03a}) with ${\bf E}$ and subtract
the results to produce a continuity equation \cite{BIMar,BIOptCommun}
\begin{equation}
\frac{1}{c}\left (
{\bf E}\cdot\frac{\partial{\bf D}}{\partial t}
+{\bf H}\cdot \frac{\partial{\bf B}}{\partial t}
\right )
+\nabla\cdot ({\bf E}\times{\bf H})=-\frac{{\bf J}_f}{c}\cdot{\bf E} 
\label{EQq2.09}
\end{equation}
that is a valid theorem (Poynting's theorem) of the formal theory of
continuum electrodynamics.
\par
Adding
the vector product of ${\bf B}$ with Eq.~(\ref{EQq2.03a}),
the vector product of ${\bf D}$ with Eq.~(\ref{EQq2.03b}),
the product of Eq.~(\ref{EQq2.03d}) with $-{\bf H}$, and
the product of Eq.~(\ref{EQq2.03c}) with $-{\bf E}$
produces the momentum continuity equation
$$
\frac{1}{c}\frac{\partial}{\partial t}({\bf D}\times{\bf B})
+{\bf D}\times(\nabla\times {\bf E})
+{\bf B}\times(\nabla\times{\bf H})
$$
\begin{equation}
-(\nabla\cdot{\bf D}){\bf E} - (\nabla\cdot{\bf B}){\bf H} =
-\rho_f{\bf E} -\frac{1}{c}{\bf J}_f\times{\bf B}
\label{EQq2.10}
\end{equation}
that is also a formal theorem of Maxwellian continuum electrodynamics
\cite{BIOptCommun}.
\par
The free charge density and the free charge current density are
parameters that are determined by the specification of the system
configuration.
Then we can specify a system that consists of a neutral 
magneto-dielectric medium situated in the vacuum and illuminated
by a finite quasimonochromatic field.
The Maxwell--Minkowski equations, Eqs.~(\ref{EQq2.03}), become 
homogeneous Maxwell--Minkowski equations,
\begin{subequations}
\begin{equation}
\nabla\times{\bf H}- \frac{\partial {\bf D}}{\partial (ct)}=0
\label{EQq2.11a}
\end{equation}
\begin{equation}
\nabla\times{\bf E}+ \frac{\partial {\bf B}}{\partial (ct)}=0
\label{EQq2.11b}
\end{equation}
\begin{equation}
\nabla\cdot{\bf D}=0
\label{EQq2.11c}
\end{equation}
\begin{equation}
\nabla\cdot{\bf B}=0
\label{EQp.11d} \;,
\end{equation}
\label{EQq2.11}
\end{subequations}
for a neutral simple linear medium in the absence of the free charge
density $\rho_f$ and the free current density ${\bf J}_f$.
\par
Reproducing the derivation of Eqs.~(\ref{EQq2.09}) and (\ref{EQq2.10})
using the homogeneous Maxwell equations Eqs.~(\ref{EQq2.11}), we
obtain the homogeneous electromagnetic continuity equations,
\begin{subequations}
\begin{equation}
\left (
{\bf E}\cdot\frac{\partial{\bf D}}{\partial (ct)}
+{\bf H}\cdot \frac{\partial{\bf B}}{\partial (ct)}
\right )
+\nabla\cdot ({\bf E}\times{\bf H})=0
\label{EQq2.12a}
\end{equation}
$$
\frac{\partial}{\partial (ct)}({\bf D}\times{\bf B})
+{\bf D}\times(\nabla\times {\bf E})
+{\bf B}\times(\nabla\times{\bf H})
$$
\begin{equation}
-(\nabla\cdot{\bf D}){\bf E} - (\nabla\cdot{\bf B}){\bf H} =0 \;,
\label{EQq2.12b}
\end{equation}
\label{EQq2.12}
\end{subequations}
that are formal theorems of the homogeneous Maxwell--Minkowski
equations for a neutral magneto-dielectric linear medium.
\par
The derivations of the electromagnetic continuity equations,
Eqs.~(\ref{EQq2.09}) and (\ref{EQq2.10}), and the homogeneous
electromagnetic continuity equations, Eqs.~(\ref{EQq2.12}),
are straightforward theorems of the Maxwell--Minkowski equations
and constitutive relations.
The derivations and results present some significant features:
\par
\textit {i}) The energy continuity equation (Poynting's theorem) and the
momentum continuity equations are identities of the Maxwell--Minkowski
equations.
The usual derivation \cite{BIBoydMil,BIJackson,BIGriff,BIZangwill}
as equations of motion of free charge density and free charge
current density, Eq.~(\ref{EQq2.05}) to Eq.~(\ref{EQq2.08}),
is not appropriate when applied to a neutral medium in which these
densities do not exist.
Therefore, the usual derivation as equations of motion of the
free charge density and free charge current density is not
appropriate, in general.
\par
\textit {ii}) The Lorentz force law is not a postulate of Maxwellian
continuum electrodynamics \cite{BIManx,BIMansurx}.
Instead, the Lorentz force density law, Eq.~(\ref{EQq2.05}), 
is a relation that is derived
as part of a theorem, Eq.~(\ref{EQq2.10}), of the macroscopic
Maxwell--Minkowski equations, Eqs.~(\ref{EQq2.03}), using the
requirement that the change in mechanical momentum is equal and
opposite to the change in electromagnetic momentum in a conservative
system.
\par
\textit {iii}) The divergence of the Poynting vector appears in the
energy continuity equations, Eq.~(\ref{EQq2.09}) and (\ref{EQq2.12a}),
so that Poynting's vector is considered arbitrary to the
extent that the curl of any vector field can be added to
it \cite{BIJackson,BIZangwill}.
Except the energy continuity equation is derived as an identity of
the Maxwell--Minkowski equations that do not admit an arbitrary
vector field in that manner.
\par
\textit {iv}) The charge continuity equation (charge conservation law)
\begin{equation}
\frac{\partial \rho_f}{\partial t}+\nabla\cdot{\bf J}_f=0
\label{EQq2.13}
\end{equation}
can be derived by substituting Eq.~(\ref{EQq2.03c}) into the
divergence of Eq.~(\ref{EQq2.03a}).
A continuity equation (conservation law), see Sec. III, describes
the unimpeded, inviscid, incoherent, incompressible flow of
non-interacting particles in the continuum limit through otherwise
empty space (vacuum).
The presence of a density of interacting charged material particles
flowing unimpeded through a continuous polarizable/magnetizable material
medium is not consistent with the conditions for a spacetime
continuity equation that is derived for noninteracting particles
in the continuum limit traveling unimpeded in the vacuum, Sec. III.
We let $\rho_f=0$ and ${\bf J}_f=0$ in order to treat the fundamental
case of propagation of the field through a neutral linear medium.
\par
\textit {v}) The theoretical procedure can be applied to derive
analogous energy and momentum equations for the microscopic fields
as identities of the microscopic Maxwell equations, instead of
as equations of motion for the free charge density and free charge
current density in the vacuum.
The comments about the continuity equations apply in similar form
to the field in the vacuum.
\par
As a matter of linear algebra, Eqs.~(\ref{EQq2.12}) can be written
row-wise as a differential equation \cite{BIOptCommun}.
We write Eq.~(\ref{EQq2.12b}) in component form as \cite{BIJackson}
$$
\frac{\partial ({\bf D}\times{\bf B})^i}{\partial (ct)}
+\sum_j\frac{\partial}{\partial x^j}{{\sf W}_c}^{ij}=
-\frac{\varepsilon{\bf E}^2}{2}\frac{\nabla\varepsilon}{\varepsilon}
-\frac{\mu{\bf H}^2}{2} \frac{\nabla \mu }{\mu}
$$
using the constitutive relations, Eqs.~(\ref{EQq2.04}).
Then \cite{BIJackson},
\begin{subequations}
\begin{equation}
\partial_{\beta}{\sf T}_{M}^{\alpha\beta}= {\bf f}_M^{\alpha}
\label{EQq2.14a}
\end{equation}
\begin{equation}
{\sf T}_{M}^{\alpha\beta} = 
\left [
\begin{matrix}
\frac{1}{2}({\bf D}\cdot{\bf E}+{\bf B}\cdot{\bf H})
&({\bf E}\times{\bf H})^1  &({\bf E}\times{\bf H})^2
&({\bf E}\times{\bf H})^3
\cr
({\bf D}\times{\bf B})^1   &{\sf W}^{11}  &{\sf W}^{12}  &{\sf W}^{13}
\cr
({\bf D}\times{\bf B})^2   &{\sf W}^{21}  &{\sf W}^{22}  &{\sf W}^{23}
\cr
({\bf D}\times{\bf B})^3   &{\sf W}^{31}  &{\sf W}^{32}  &{\sf W}^{33}
\cr
\end{matrix}
\right ]
\label{EQq2.14b}
\end{equation}
\begin{equation}
{\sf W}^{ij}= -D^iE^j-B^iH^j+
\frac{1}{2}({\bf D}\cdot{\bf E}+{\bf B}\cdot{\bf H})\delta^{ij}
\label{EQq2.14c}
\end{equation}
\begin{equation}
{\bf f}^{\alpha}_M=
\left (0,-\frac{\varepsilon{\bf E}^2}{2}\frac{\nabla\varepsilon}
{\varepsilon}
-\frac{\mu{\bf H}^2}{2}
\frac{\nabla \mu }{\mu}\right ) 
\label{EQq2.14d}
\end{equation}
\begin{equation}
\partial_{\beta}=\left (
\frac{\partial}{\partial (ct)},
\frac{\partial}{\partial x},
\frac{\partial}{\partial y},
\frac{\partial}{\partial z}
 \right ) 
\label{EQq2.14e}
\end{equation}
\label{EQq2.14}
\end{subequations}
is a formal theorem of the homogeneous electromagnetic continuity
equations, Eqs.~(\ref{EQq2.12}), as well as a formal theorem of the
homogeneous Maxwell--Minkowski equations, Eqs.~(\ref{EQq2.11}).
Specifically, the Minkowski energy--momentum tensor (matrix)
${\sf T}_{M}^{\alpha\beta}$, Eq.~(\ref{EQq2.14b}), the 
Minkowski energy density $u={\sf T}_{M}^{00}$, and the
Minkowski momentum density
${\bf g}_M=({\sf T}_{M}^{10},{\sf T}_{M}^{20},{\sf T}_{M}^{30})$
are formally derived from the axioms of continuum electrodynamics,
the Maxwell--Minkowski and constitutive equations, as part of the
theorem for the continuity equation, Eq.~(\ref{EQq2.14a}).
\par
In the limit that the gradient Minkowski four-force density
${\bf f}^{\alpha}_{M}$ is negligible, Eq.~(\ref{EQq2.14a}) becomes
\begin{equation}
\partial_{\beta}{\sf T}_{M}^{\alpha\beta}= 0 \;,
\label{EQq2.15}
\end{equation}
which is known as the electromagnetic conservation law.
An equivalent statement is that the Minkowski momentum is
`almost' conserved based on the identity, Eq.~(\ref{EQq2.14}), in
the case the Minkowski four-force can be treated as negligible
\cite{BIPfei,BIBrevCons,BIBrevVec,BIBrevik,BIWang4Vec}.
Conservation of the Minkowski energy and Minkowski momentum
is an obviously correct implementation of the electromagnetic
conservation law, Eq.~(\ref{EQq2.15}), and there is a large body
of work that is based on conservation of the Minkowski
momentum \cite{BIPfei,BIBrevCons,BIBrevVec,BIBrevik,BIWang4Vec}.
\par
In contradiction, there is a large body of scientific work that
proves that the Minkowski momentum is neither conserved nor almost
conserved \cite{BIPfei,BIFofn1,BIFofn2,BIGord}.
The wave equation 
\begin{equation}
\nabla\times(\nabla\times{\bf A})+
\frac{n^2}{c^2}\frac{\partial^2 {\bf A}}{\partial t^2} 
=\frac{\nabla\mu}{\mu}\times(\nabla\times{\bf A})
\label{EQq2.16}
\end{equation}
is also a theorem of the Maxwell--Minkowski equations, 
Eq.~(\ref{EQq2.11}),
with the constitutive relations, Eq.~(\ref{EQq2.04}),
and the Coulomb-gauge definition of the macroscopic fields 
\begin{subequations}
\begin{equation}
{\bf E}=- \frac{\partial {\bf A}}{\partial (ct)}
\label{EQq2.17a}
\end{equation}
\begin{equation}
{\bf B}=\nabla\times {\bf A}
\label{EQq2.17b}
\end{equation}
\label{EQq2.17}
\end{subequations}
in terms of the vector potential ${\bf A}$.
The Coulomb gauge is suitable for a sourceless medium, $\rho_f=0$
and ${\bf J}_f=0$, allowing the scalar potential $\Phi$ to be
suppressed.
\par
The derivation of the wave equation theorem, Eq.~(\ref{EQq2.16}),
consists of substituting Eqs.~(\ref{EQq2.04}) and (\ref{EQq2.17})
into the homogeneous Maxwell--Amp\`ere law, Eq.~(\ref{EQq2.11a}).
Repeated analyses of the wave equation and wave propagation, for
over a century, have disclosed that the Minkowski electromagnetic
momentum in an antireflection-coated transparent linear dielectric
is greater that the incident momentum by a non-negligible
multiplicative factor of $n$ \cite{BIPfei,BIFofn1,BIFofn2,BIGord}.
Acknowledgment of this easily verified theoretical fact is present 
in the scientific record, where the Minkowski
pull-force is the hypothetical source of
this momentum difference and there is no need to repeat the
wave propagation analyses here.
\par
In order to be complete, but concise, we provide a short demonstration 
using global conservation of energy to prove that Minkowski linear
momentum is not conserved in a linear dielectric
\cite{BIPfei,BIFofn1,BIFofn2,BIGord}.
For a monochromatic field of frequency $\omega_f$ with
refractive index $n(\omega_f)$ in which the vector
potential amplitude of the incident field is $A_0$, the Minkowski
energy density \cite{BIJackson,BIGriff,BIZangwill} of the field
in the medium is
\begin{equation}
u_M=\frac{1}{2}({\bf D}\cdot{\bf E}+{\bf B}\cdot{\bf B}) 
=\frac{\omega_f^2 n^2}{2c^2}|A|^2
=\frac{\omega_f^2 n}{2c^2}|A_0|^2
\label{EQq2.18}
\end{equation}
in the plane-wave limit.
\par
Due to the reduced velocity of light in the dielectric,
a quasimonochromatic field in the plane-wave limit has an
extent along the propagation direction (the longitudinal width)
that differs from the longitudinal extent of the incident field
$w$ by a factor of $1/n$ 
\cite{BIOptCommun}.
The Minkowski energy of a
quasimonochromatic field of cross-sectional area $\sigma$
\begin{equation}
U_M=\frac{1}{2}\int_{\Sigma}
({\bf D}\cdot{\bf E}+{\bf B}\cdot{\bf B})\sigma dz 
=\frac{\omega_f^2 w \sigma}{2c^2}|A_0|^2
\label{EQq2.19}
\end{equation}
is constant in time as the field propagates from the
vacuum (longitudinal field width $w$) and into the dielectric
(width $w/n$) through a gradient-index antireflection coating
in the plane-wave limit.
For the same quasimonochromatic field, the Minkowski momentum is
\begin{equation}
{\bf G}_M=\int_{\Sigma} \frac{{\bf D}\times{\bf B}}{c}
\sigma dz 
=\frac{\omega_f^2 n w \sigma}{2c^2}|A_0^2|{\bf \hat k}
\label{EQq2.20}
\end{equation}
based on the Minkowski momentum density
\begin{equation}
{\bf g}_M=\frac{{\bf D}\times{\bf B} }{c}
=\frac{\omega_f^2 n^2}{2c^2}|A_0^2|{\bf \hat k} \, .
\label{EQq2.21}
\end{equation}
Comparing the formula for the Minkowski momentum,
Eq.~(\ref{EQq2.20}), with the formula for the conserved energy,
Eq.~(\ref{EQq2.19}), on the basis of the vector potential magnitude 
shows that the momentum of the electromagnetic field in the medium is
not globally conserved by a factor of $n$ for a finite field, even
though this contradicts the (local) electromagnetic conservation law,
Eq.~(\ref{EQq2.15}).
\textit{The fact that a theorem of Maxwellian
continuum electrodynamics is proven false by another theorem of
Maxwellian continuum electrodynamics proves that one or more of
the axioms of the formal theory, the Maxwell--Minkowski equations,
the constitutive relations, and the vector potential relations,
are false.}
\par
Incomplete is also false, nuanced false, but false nevertheless.
The extant resolution of the Abraham--Minkowski controversy consists
of adding a phenomenological material-motion energy--momentum tensor
to a Maxwellian electromagnetic energy--momentum tensor (or adding
a phenomenological material-motion momentum to a Maxwellian
electromagnetic momentum)
\cite{BIRL,BIBoydMil,BIPfei,BIAMC4,BIKemplatest,BIPenHaus,BIGord,BIBarn,BIBarnLou,BIKranys,BIObukPLA,BIBrevCons}.
The resolution is a tautology: the whole is the sum of the parts.
However, the Maxwellian electromagnetic subsystem is still incomplete
because the Maxwell--Minkowski equations are not coupled to the
material subsystem equations of motion.
Likewise, the material equations of motion remain incomplete.
Instead of completing the subsystem equations of motion for both
subsystems, the electrodynamic energy--momentum tensor is
superficially coupled to the energy--momentum tensor for the
material through the transient force term, ${\bf f}^{\alpha}_M$,
of an arbitrarily long field.
\par
The medium is typically modeled as dust \cite{BIPfei}, an 
unimpeded, inviscid, incoherent, incompressible flow of
non-interacting particles of mass-bearing matter in the continuum
limit through empty space.
The total energy and total momentum are known quantities because the
energy and momentum of the incident field are known.
Then conservation of total energy and conservation of total momentum
are used to derive the adjustable material parameters, the particle
density and velocity \cite{BIPfei}.
We will see below that a microscopic model of the medium is not
required because the total energy and total momentum are known
by global conservation because the incident energy and the
incident momentum are specified.
\par
The material-motion momentum that supplements the Minkowski
electromagnetic momentum is identified by
Barnett and Loudon \cite{BIBarnLou} as the material canonical
momentum ${\bf G}^{medium}_{canonical}$ such that
\begin{equation}
{\bf G}_{tot}=
{\bf G}_M+
{\bf G}^{medium}_{canonical} 
\label{EQq2.22}
\end{equation}
is the total momentum ${\bf G}_{tot}$.
In the context of continuum electrodynamics, whatever microstructure
of the material and field that exists in nature is treated in the
continuum limit so that only the continuous linear response is left.
Then, the particular microscopic model of the linear medium cannot
matter and the material canonical momentum is given as
${\bf G}^{medium}_{canonical}={\bf G}_{tot}-{\bf G}_M$,
where the total momentum ${\bf G}_{tot}$, the Minkowski momentum
${\bf G}_M$, and the material canonical momentum
${\bf G}^{medium}_{canonical}$ are all macroscopic quantities
and are continuous at all length scales ($\sum_n^N \rightarrow \int dv$)
in the continuum limit.
\par
Using global conservation of total momentum in a closed
system produces formulas for a total (field plus material)
momentum \cite{BIGord}
\begin{equation}
{\bf G}_{tot}=\int_{\Sigma} \frac{n{\bf E}\times{\bf B}}{c}
\sigma dz 
\label{EQq2.23}
\end{equation}
and a material canonical momentum
\begin{equation}
{\bf G}^{medium}_{canonical}=
\int_{\Sigma} \frac{(n-n^2){\bf E}\times{\bf B}}{c}
\sigma dz 
\label{EQq2.24}
\end{equation}
based on the momentum of the incident field.
The total (field plus material) momentum ${\bf G}_{tot}$ was
constructed by Gordon \cite{BIGord} to be constant in time for
the field in a dielectric.
However, Gordon uses the concept of pseudomomentum to reintroduce
the extra factor of $n$ in the total momentum.
closed system and the Gordon total momentum successfully addresses
the factor of $n$ error in global conservation of linear momentum.
\par
The consensus resolution of the Abraham--Minkowski controversy
is circular, accomplishing global conservation of linear momentum
by fiat.
A circular theory proves itself in the context in which it was
derived.
The total linear momentum, Eq.~(\ref{EQq2.23}), that comes out of
the system of subsystems treatment is provably correct because
it was derived by global conservation principles \cite{BIGord}.
For a linear dielectric medium, the penultimate result of the system
of subsystems approach is the total (field plus material)
energy--momentum tensor \cite{BIPfei,BIJMP}
\begin{subequations}
$$
{\sf T}_{tot}^{\alpha\beta} = 
$$
\begin{equation}
\left [
\begin{matrix}
\frac{1}{2}(n^2{\bf E}\cdot{\bf E}+{\bf B}\cdot{\bf B})
&(n{\bf E}\times{\bf B})^1 &(n{\bf E}\times{\bf B})^2
&(n{\bf E}\times{\bf B})^3
\cr
(n{\bf E}\times{\bf B})^1   &{\sf W}_{tot}^{11}  &{\sf W}_{tot}^{12}  &{\sf W}_{tot}^{13}
\cr
(n{\bf E}\times{\bf B})^2   &{\sf W}_{tot}^{21}  &{\sf W}_{tot}^{22}  &{\sf W}_{tot}^{23}
\cr
(n{\bf E}\times{\bf B})^3   &{\sf W}_{tot}^{31}  &{\sf W}_{tot}^{32}  &{\sf W}_{tot}^{33}
\cr
\end{matrix}
\right ] 
\label{EQq2.25a}
\end{equation}
\begin{equation}
{\sf W}_{tot}^{ij}= -n^2E^iE^j-B^iB^j+
\frac{1}{2}(n^2{\bf E}\cdot{\bf E}+{\bf B}\cdot{\bf B})\delta^{ij} \, .
\label{EQq2.25b}
\end{equation}
\label{EQq2.25}
\end{subequations}
The total energy $U_{tot}=\int_{\Sigma} {\sf T}_{tot}^{00}\sigma dz$
and the total momentum
${\bf G}_{tot}=
\int_{\Sigma} ({\sf T}_{tot}^{01},{\sf T}_{tot}^{02},
{\sf T}_{tot}^{03})\sigma dz$
are demonstrably constant in time for our model system.
However, substituting the total energy--momentum
tensor, Eq.~(\ref{EQq2.25a}), 
into the local electromagnetic conservation
law (see Eq.~(\ref{EQq3.01}) and compare Eq.~(\ref{EQq2.15}))
\begin{equation}
\partial_{\beta}{\sf T}_{tot}^{\alpha\beta}=0 \;,
\label{EQq2.26}
\end{equation}
one obtains
\begin{equation}
\frac{\partial}{\partial (ct)}
\left [\frac{1}{2} (\varepsilon{\bf E}\cdot{\bf E}
+{\bf B}\cdot{\bf B})\right ]+
\nabla \cdot(n{\bf E}\times{\bf B})=0
\label{EQq2.27}
\end{equation}
for the $\alpha=0$ component.
Equation (\ref{EQq2.27}) violates Poynting's theorem and the 
equation is self-inconsistent because the non-zero terms depend
on different powers of $n$.
Then the consensus resolution of the Abraham--Minkowski controversy
in terms of the total (field plus material) energy--momentum tensor
(or the total (field plus material) momentum) is demonstrably false,
even though important portions have been proven true.
\par
The material subsystem conjecture has been disproved by showing
that the total (field plus material) energy--momentum tensor that heals
the violation of the global conservation law introduces violations of
the spacetime (local) conservation law (including Poynting's
theorem).
Although cast in terms of the Minkowski energy--momentum tensor,
the disproof works equally well with the Abraham energy--momentum
tensor because the total (field plus material) energy--momentum tensor
${\sf T}_{tot}^{\alpha\beta}$, Eq.~(\ref{EQq2.25a}), is the same
in both cases \cite{BIPfei}.
\par
Because the Maxwell--Minkowski model is assumed to be incomplete,
one can propose other physically motivated subsystems in an attempt
to resolve the conservation issue.
Dispersion has been suggested and phenomenologically added to the
theoretical model \cite{BIBoydMil}.
The way our system is defined includes dispersion to lowest order
so the inclusion of additional dispersion is an exercise in
complexity for a second-order consequence.
Because the total energy and the total momentum, including
dispersion, are globally conserved, the total energy--momentum tensor
remains given by Eq.~(\ref{EQq2.25a}), violating self-consistency,
the Poynting theorem, and the local electromagnetic conservation
law.
\par
We can identify other inconsistent physics in the formal theory of
continuum electrodynamics.
In Ref.~\cite{BIIdentity}, the set of macroscopic field equations,
\begin{subequations}
\begin{equation}
\frac{\nabla}{n_m}\times(n_m{\bf H})
+\frac{\partial(-n_e{\bf E})}{\partial (ct/n_e)}
=\frac{\nabla n_m}{n_mn_m}\times(n_m{\bf H})
\label{EQq2.28a}
\end{equation}
\begin{equation}
\frac{\nabla}{n_m}\cdot(n_m{\bf H})
=\frac{\nabla n_m}{n_mn_m}\cdot(n_m{\bf H})
\label{EQq2.28b}
\end{equation}
\begin{equation}
\frac{\nabla}{n_m}\times(-n_e{\bf E})
-\frac{\partial(n_m{\bf H})}{\partial (ct/n_e)}
 =\frac{\nabla n_e}{n_mn_e}\times(-n_e{\bf E})
\label{EQq2.28c}
\end{equation}
\begin{equation}
\frac{\nabla}{n_m}\cdot(-n_e{\bf E})
=-\frac{\nabla n_e}{n_mn_e}\cdot(-n_e{\bf E})
\label{EQq2.28d}
\end{equation}
\begin{equation}
n_e=\sqrt{\varepsilon}
\label{EQq2.28e}
\end{equation}
\begin{equation}
n_m=\sqrt{\mu}
\label{EQq2.28f}
\end{equation}
\label{EQq2.28}
\end{subequations}
was rigorously derived as an identity of the homogeneous
Maxwell--Minkowski equations, Eqs.~(\ref{EQq2.11}), with
constitutive relations, Eqs.~(\ref{EQq2.04}), for a simple
linear medium with macroscopic fields $-n_e{\bf E}$ and $n_m{\bf H}$.
The derivation \cite{BIIdentity} is simple, reproducible, and correct.
\par
Equations (\ref{EQq2.28a})--(\ref{EQq2.28d}) are isomorphic to the
vacuum Maxwell field equations with a timelike coordinate of $ct/n_e$,
instead of $ct$, and spatial coordinates $(n_mx,n_my,n_mz)$ in the
limit that the gradients of the permittivity and permeability may
be neglected.
Then, Eqs.~(\ref{EQq2.28}) are inconsistent with Laue's implementation
of Einstein's relativity in a continuous linear medium \cite{BILaue}.
Clearly, there is an existential inconsistency associated with
Eqs.~(\ref{EQq2.28}) and (\ref{EQq2.11}) because a simple
application of algebra and calculus changes the new field equations
back to the Maxwell--Minkowski equations and the two expressions
of the identity correspond to two different relativities
with different timelike coordinates, $x_0=ct$ and $\bar x_0=ct/n_e$.
The Maxwell--Minkowski equations, Eqs.~(\ref{EQq2.03})--(\ref{EQq2.04})
and (\ref{EQq2.28}), are proven false by contradiction.
\par
The century-old momentum contradiction at the center of continuum
electrodynamics stands very much unresolved.
Moreover, issues with Maxwellian continuum electrodynamics now
extend beyond angular momentum conservation and global linear
momentum conservation to include consistency with the local
energy conservation law, Poynting's theorem, and
special relativity in a linear medium.
The formal equivalence of incommensurate macroscopic field equations,
Eqs.~(\ref{EQq2.11}) and Eqs.~(\ref{EQq2.28}), proves that the axioms
of continuum electrodynamics, the Maxwell--Minkowski and constitutive
equations, are manifestly false.
\par
Einstein taught that fundamental physical principles are rooted in 
the vacuum.
The vacuum was later formalized as an isotropic, homogeneous, flat,
four-dimensional, Minkowski spacetime ${\sf S}_M(ct,x,y,z)$.
The microscopic Maxwellian model of a linear medium consists of
tiny bits of matter embedded in the vacuum with the permittivity
$\varepsilon=1+\chi_e$ and the permeability $\mu=1+\chi_m$ defined
in terms of the unit vacuum electric susceptibility,
the unit vacuum magnetic susceptibility,
the material electric susceptibility $\chi_e$,
and the material magnetic susceptibility $\chi_m$.
As long as the individual particles of the medium are localized and
the interactions of each particle with the microscopic field are
perturbative, the flat, four-dimensional, empty Minkowski spacetime
is ``regarded as the proper setting within which to formulate those
laws of physics that do not refer specifically to gravitational
phenomena'' \cite{BINabor}.
\par
Optically transparent material are mostly empty space in which
light travels at an instantaneous speed of $c$ \cite{BIBorn}.
The tiny polarizable and magnetizable bits of matter that are embedded
in the vacuum scatter and delay the light.
Even if one intends to build a microscopic model of physical optics in
Minkowski spacetime, there are far to many particles and far too many
interactions to keep track of in `real' materials.
Consequently, in continuum electrodynamics, the phenomenological model
of the medium is an abstraction that is continuous at all length
scales from the very outset and the effective speed of light is $c/n$.
The interstitial vacuum has no role in the continuum limit and
a continuous medium with a macroscopic refractive index $n$ cannot
be re-discretized or un-averaged.
\par
In this article, we use field theory to derive equations of motion 
for electromagnetic fields in continuous linear materials starting
from identifiable and characterizable principles.
We show that an isotropic, homogeneous, flat, four-dimensional,
continuous non-Minkowski `material' spacetime is the proper
setting for continuum electrodynamics, conservation laws,
special relativity, invariance, and other optical principles
that take place in an isotropic, homogeneous, magneto-dielectric
linear medium in which the effective speed of light is $c/n$.
Each different isotropic, homogeneous, transparent, linear medium will
be associated with a different continuous `material' spacetime
connected to other spacetimes by boundary conditions.
\par
\section{Spacetime Conservation Laws}
\par
Special relativity, Laue's theorem \cite{BIGiu,BIWang2}, and Noether's
theorem \cite{BINoether} constitute a powerful framework within which
 to analyze energy and momentum conservation of a continuous flow of
light.
In fact, so much of the physics is performed by the formalism that
our problem with conservation of momentum in a simple linear medium
is embedded in the re-application of the relativistic formalism of
physics in a vacuum to a continuous medium.
\par
The tensor energy--momentum formalism is well-known when applied to
continuum (fluid) dynamics.
Before treating conservation laws in a linear medium, we review what
is typically known about conservation laws in the vacuum
of an otherwise empty Minkowski spacetime.
\par
\textit{a}) For a thermodynamically closed system, the local spacetime
conservation law of the total system
\begin{equation}
\partial_{\beta}{\sf T}_{tot}^{\alpha\beta}=0
\label{EQq3.01}
\end{equation}
is derived by applying the divergence theorem to a Taylor series
expansion of the density field of the energy and momentum properties
of an unimpeded, inviscid, incoherent, incompressible, flow of
non-interacting particles (nonpolar fluid molecules, dust particles,
or photons) in the continuum limit (fluid or light field
(light fluid)) in an \textit{otherwise empty volume}
(vacuum) \cite{BIFox}.
The local spacetime conservation law, Eq.~(\ref{EQq3.01}), is a
theorem of the field theory and is characteristic of a 
conserved flow in Minkowski spacetime ${\sf S}_M(ct,x,y,z)$.
The four-divergence of the energy--momentum tensor must vanish
as a condition for conservation of an unimpeded, inviscid,
incoherent, incompressible flow of non-interacting particles in the
continuum limit through empty space \cite{BIFox,BIGold}.
\par
\textit{b}) Under typical conditions, the energy density and
the momentum density integrated over the total volume $\Sigma$ of the
thermodynamically closed system
\begin{equation}
U=\int_{\Sigma} {\sf T}_{tot}^{00} dv
\label{EQq3.02}
\end{equation}
\begin{equation}
{\bf G}=\left (\frac{1}{c} \int_{\Sigma}{\sf T}_{tot}^{01} dv,
\frac{1}{c}\int_{\Sigma}{\sf T}_{tot}^{02} dv,
\frac{1}{c}\int_{\Sigma}{\sf T}_{tot}^{03} dv \right )
\label{EQq3.03}
\end{equation}
must be constant in time (global conservation).
The system can be as large as is required to completely contain
the matter and energy, but the boundaries of the closed
system will still be definite (arbitrarily large).
The conservation conditions, Eqs.~(\ref{EQq3.02}) and (\ref{EQq3.03}),
require no matter or energy crossing the boundary of the system as an
initial condition $(-\infty < t_0 \leq t)$.
(Zero-field boundary conditions for all time $(-\infty <t< \infty)$
correspond to an empty or static system \cite{BIWang2}).
Examples of non-conservative systems for which the global conservation
laws, Eqs.~(\ref{EQq3.02}) and (\ref{EQq3.03}), fail include systems
in which a source or sink is present, unbounded systems, subsystems
of a complete system, and inconsistently defined systems.
\par
\textit{c}) For typical conditions in which the energy--momentum
tensor of the initial flow is diagonally symmetric,
or is transformed into a symmetric tensor, the energy--momentum
tensor of a closed system must remain symmetric
\begin{equation}
{\sf T}_{tot}^{\alpha\beta}={\sf T}_{tot}^{\beta\alpha}
\label{EQq3.04}
\end{equation}
in order to conserve angular momentum.
This condition explicitly couples the rows of the energy--momentum
tensor.
Obviously, if the incident field contains angular momentum then the
energy--momentum tensor of a conservative system will not be symmetric.
\par
It is possible to write, pro forma, a matrix-based differential
equation from continuity equations of different systems or subsystems.
Such a compound system is inconsistent and that is discovered 
by the non-symmetric matrix that results from the lack of coupling
between the continuity equations.
Pathological exceptions to symmetry may include non-symmetric initial
and boundary conditions, unclosed systems (subsystems), inhomogeneous
systems that include microstructure, non-isotropic systems,
coordinate system changes, and inconsistently defined systems.
Pathological conditions are not likely in the middle of free space,
but the issue is presaged for the case of propagation of light from
the vacuum into a simple linear medium where the non-symmetric
Minkowski energy--momentum tensor has come to be viewed as acceptable.
\par
\textit{d}) The trace of the energy--momentum tensor is the density
of the fluid
\begin{equation}
\rho=g_{\alpha\alpha}{\sf T}_{tot}^{\alpha\alpha}
\label{EQq3.05}
\end{equation}
with metric tensor $g_{\alpha\beta}$ for a non-pathological closed system.
\par
\textit{e}) The local conservation law is sometimes written as \cite{BIRL,BIPfei}
\begin{equation}
\partial_{\alpha}{\sf T}_{tot}^{\alpha\beta}=0 \;.
\label{EQq3.06}
\end{equation}
This condition is typically true for a conserved system, derived by
substituting Eq.~(\ref{EQq3.04}) into Eq.~(\ref{EQq3.01}).
However, Eq.~(\ref{EQq3.06}) cannot be considered
a conservation law in the sense of Eq.~(\ref{EQq3.01}) because it
implicitly includes an additional condition, namely diagonal symmetry
of the energy--momentum tensor.
\par
The conservation law, Eq.~(\ref{EQq3.01}), is derived \cite{BIFox}
using spacetime coordinates $(ct,x,y,z)$ and it is \textit {manifestly}
\textbf{not} \textit{dependent on the Maxwell field equations}.
To be sure, the energy and momentum of an inviscid, incoherent,
incompressible flow of non-interacting photons propagating unimpeded
in the vacuum are conserved and therefore must be consistent with
the spacetime conservation laws, Eqs.~(\ref{EQq3.01})--(\ref{EQq3.05}).
Now,
\begin{subequations}
\begin{equation}
\partial_{\beta}{\sf T}_{vac}^{\alpha\beta}=0
\label{EQq3.07a}
\end{equation}
\begin{equation}
{\sf T}_{vac}^{\alpha\beta} = 
\left [
\begin{matrix}
\frac{1}{2}({\bf E}\cdot{\bf E}+{\bf B}\cdot{\bf B})
&({\bf E}\times{\bf B})^1  &({\bf E}\times{\bf B})^2
&({\bf E}\times{\bf B})^3
\cr
({\bf E}\times{\bf B})^1   &{\sf W}_{vac}^{11}  &{\sf W}_{vac}^{12}  &{\sf W}_{vac}^{13}
\cr
({\bf E}\times{\bf B})^2   &{\sf W}_{vac}^{21}  &{\sf W}_{vac}^{22}  &{\sf W}_{vac}^{23}
\cr
({\bf E}\times{\bf B})^3   &{\sf W}_{vac}^{31}  &{\sf W}_{vac}^{32}  &{\sf W}_{vac}^{33}
\cr
\end{matrix}
\right ] 
\label{EQq3.07b}
\end{equation}
\begin{equation}
{\sf W}_{vac}^{ij}= -E^iE^j-B^iB^j+
\frac{1}{2}({\bf E}\cdot{\bf E}+{\bf B}\cdot{\bf B})\delta^{ij} \, .
\label{EQq3.07c}
\end{equation}
\label{EQq3.07}
\end{subequations}
is a theorem of the energy and momentum continuity equations that are
typically derived in electricity and magnetism/electrodynamics
textbooks using the microscopic Maxwell equations for light fields
in the vacuum \cite{BIGriff}.
Then Eq.~(\ref{EQq3.07a}) is considered to be the electromagnetic
conservation law based on the fact that Eq.~(\ref{EQq3.07}) is a
theorem of the fundamental (vacuum) Maxwell equations of
electrodynamics, the similar appearance of Eqs.~(\ref{EQq3.07a})
and (\ref{EQq3.01}), and a physical necessity argument that a
closed system consisting of a finite quasimonochromatic field
propagating in the vacuum is conserved.
\par
However, the principles of conservation are nowhere used in the
derivation of Eq.~(\ref{EQq3.07}) from the microscopic Maxwell
equations and several important conditions are not incorporated
into the derivation of Eq.~(\ref{EQq3.07}).
Therefore it is not strictly correct to identify Eq.~(\ref{EQq3.07})
as `the' electromagnetic conservation law unless the closed system
satisfies all conservation laws, Eqs.~(\ref{EQq3.01})--(\ref{EQq3.05}),
zero-field boundary conditions with the entire field contained within
the boundaries of the system at a finite time in the past, and the
predicate of unimpeded, inviscid, incoherent, incompressible flow of
non-interacting photons in the continuum limit through empty space.
\par
The spacetime conservation laws, Eqs.~(\ref{EQq3.01})--(\ref{EQq3.05}),
are satisfied by a quasimonochromatic field propagating in the 
vacuum of free space in the plane-wave limit.
This is easily demonstrated by substituting the elements of the 
vacuum-based energy--momentum tensor, Eq.~(\ref{EQq3.07b}),
into the conservation laws, Eqs.~(\ref{EQq3.01})--(\ref{EQq3.05}),
with $g_{\alpha\beta}= {\rm diag}(1,-1,-1,-1)$.
Condition Eq.~(\ref{EQq3.05}) shows that the trace of the
energy--momentum tensor is zero corresponding to massless photons.
Then Eq.~(\ref{EQq3.07}) can indeed be considered to be the 
spacetime conservation law for the electrodynamics of fields
in the vacuum.
\par
Next, we switch from the propagation of electromagnetic fields
in the vacuum to propagation in a linear medium.
Consider the application of the spacetime conservation laws,
Eqs.~(\ref{EQq3.01})--(\ref{EQq3.05}),
to the propagation of a continuous light field in a continuous
linear medium.
Substituting elements of the Minkowski energy--momentum tensor,
Eq.~(\ref{EQq2.14b}), into the spacetime conservation laws, we
find that the global momentum, Eq.~(\ref{EQq3.03}), is not
constant in time and that the symmetry law, Eq.~(\ref{EQq3.04}),
is violated, as expected based on the discussion in Sec. I.
The recognized fix is to use global conservation to supplement
the macroscopic Minkowski energy--momentum tensor with a
phenomenological material motion energy--momentum tensor to
create a total, field plus matter, energy--momentum tensor.
Substituting elements of the total, field plus matter, energy--momentum
tensor, Eq.~(\ref{EQq2.25a}), into the conservation laws, we find
that the $\alpha=0$ element of the local spacetime conservation
law, Eq.~(\ref{EQq3.01}), reproduces Eq.~(\ref{EQq2.27}) that is
self-inconsistent and violates Poynting's theorem.
The local conservation law and the global conservation law are
inconsistent in this case because a continuous linear medium does
not meet the condition of an otherwise empty volume for the
application of the conservation laws.
\par
\section{Lagrangian Density}
\par
Substituting the elements of the macroscopic Minkowski energy--momentum
tensor, which is derived as a theorem from the Maxwell--Minkowski
equations, into the spacetime conservation laws
Eqs.~(\ref{EQq3.01})--(\ref{EQq3.05}), proves that the macroscopic
system violates conservation of angular momentum and
violates conservation of linear momentum.
Rather than start anew, the accepted approach has been to treat the
system as incomplete and propose supplemental energy--momentum tensors.
The complete energy--momentum tensor is known by using global
conservation of energy and momentum to derive the necessary elements
of the total energy--momentum tensor \cite{BIPfei,BIJMP,BIOptCommun}.
Substituting these elements into the local conservation law produces
a false statement, Eq.~(\ref{EQq2.27}).
Then the Maxwell--Minkowski equations are manifestly false and the 
equations of motion of the total (field plus material) system are
also false.
Having proven the existing macroscopic theory to be false, we are
starting with a clean slate for the construction of an entirely new
formalism of continuum electrodynamics.
\par
Theoretical physics in a simple linear medium that is continuous
at all length scales is a problem that is multiply connected
with a large variety of places to start.
But if we enforce consistency at the boundaries between
electrodynamics, relativity, invariance, spacetime, electromagnetic
boundary conditions, etc, then we should arrive at the same set of
results no matter where we start.
\par
Lagrangian field theory is a generalization of particle dynamics to a
continuous field \cite{BIGold,BICohen}.
The classical Lagrangian is 
\begin{equation}
L=\frac{1}{2} \int_{\Sigma} (T-V) dv \, ,
\label{EQq4.01}
\end{equation}
where $T$ is the kinetic energy density, $V$ is the potential energy
density, and integration is performed over a closed system $\Sigma$.
For the electromagnetic field in a source-free simple linear medium,
the classical Lagrangian, Eq.~(\ref{EQq4.01}), can be written as
\begin{equation}
L=\frac{1}{2}\int_{\Sigma}
\left ({\varepsilon}{\bf E}^2-{\bf B}^2/\mu\right )dv
\label{EQq4.02}
\end{equation}
in the common Maxwell--Minkowski formulation of Maxwellian continuum
electrodynamics \cite{BIGold,BICohen,BIJackson,BIZangwill}.
The corresponding Lagrangian density is
\begin{equation}
{\cal L}=\frac{1}{2}
\left ({\varepsilon}{\bf E}^2-{\bf B}^2/\mu\right )
=\frac{1}{2} \left (
\varepsilon\left (
\frac{\partial{\bf A}}{\partial (ct)}
\right )^2
- \frac{ \left (\nabla\times{\bf A} \right )^2 }{\mu} 
\right )\, .
\label{EQq4.03}
\end{equation}
Now, we can use \cite{BINewFres,BIIdentity}
\begin{equation}
n_e=\sqrt{\varepsilon}
\label{EQq4.04}
\end{equation}
to denote the electric component of the refractive index and the
magnetic component of the refractive index can be denoted by
\begin{equation}
n_m=\sqrt{\mu} \; .
\label{EQq4.05}
\end{equation}
\par
The electric refractive index $n_e$, like the electric permittivity
$\varepsilon$, is clearly associated with the kinetic energy density
$T$ of the Lagrangian.
The magnetic refractive index $n_m$ and the magnetic permeability $\mu$ 
are clearly associated with the potential energy density component $V$
of the Lagrangian.
Using simple algebra, the classical Lagrangian, Eq.~(\ref{EQq4.02}),
can be written as
\begin{equation}
L=\frac{1}{2} \int_{\Sigma} \left (
\left (
\frac{n_e}{c}\frac{\partial{\bf A}}{\partial t}
\right )^2
- \left (\frac{\nabla\times{\bf A}}{n_m} \right )^2
\right ) dv \, .
\label{EQq4.06}
\end{equation}
The Lagrangian density,
\begin{equation}
{\cal L}=\frac{1}{2}\left (
\left (
\frac{n_e}{c}\frac{\partial{\bf A}}{\partial t}
\right )^2
- \left (\frac{\nabla\times{\bf A}}{n_m} \right )^2
\right ) \, ,
\label{EQq4.07}
\end{equation}
is the integrand of the Lagrangian, Eq.~(\ref{EQq4.06}).
\par
We consider an arbitrarily large region of space to be filled with an
isotropic, homogeneous, transparent, continuous, linear
magneto-dielectric medium that can be characterized by a macroscopic
electric refractive index $n_e$ and a macroscopic magnetic refractive
index $n_m$.
Treating dispersion in lowest order, the electric and magnetic
refractive indices will depend on the center frequency of the
quasimonochromatic field (or the frequency of a monochromatic field)
that illuminates the medium. 
\par
We limit our attention to an arbitrarily large simple linear medium and
we write a new time-like variable
\begin{equation}
\bar x^0=\frac{ct}{n_e}
\label{EQq4.08}
\end{equation}
and new spatial variables
\begin{subequations}
\begin{equation}
\bar x=n_m x
\label{EQq4.09a}
\end{equation}
\begin{equation}
\bar y=n_m y
\label{EQq4.09b}
\end{equation}
\begin{equation}
\bar z=n_m z 
\label{EQq4.09c}
\end{equation}
\label{EQq4.09}
\end{subequations}
based on the way the electric and magnetic indices of refraction 
appear in the Lagrangian density.
Although we can retain the spatial and temporal dependencies of the
components of the refractive index, in this work we have adopted the 
limit of an isotropic homogeneous medium in which these dependence's can
be neglected in order to proceed with the fundamental physical issues.
As always, we can treat a piecewise homogeneous medium by using the
homogeneous theory plus boundary conditions.
Further, we have specified conditions that allow dispersion to be
treated to lowest order and velocity-dependent anisotropy to be
neglected.
\par
We construct a `material' Laplacian operator 
\begin{equation}
\bar \nabla = 
\left ( \frac{\partial}{\partial \bar x},
\frac{\partial}{\partial \bar y},
\frac{\partial}{\partial \bar z} \right ) 
\label{EQq4.10}
\end{equation}
to be used in the abstract mathematical model of an arbitrarily large,
isotropic, homogeneous simple linear medium.
Substituting Eqs.~(\ref{EQq4.08})--(\ref{EQq4.10}) into
Eqs.~(\ref{EQq4.06}) and (\ref{EQq4.07}), we obtain a Lagrangian,
\begin{equation}
L=\frac{1}{2} \int_{\Sigma} 
\left ( \left (
\frac{\partial{\bf A}}{\partial \bar x^0}
\right )^2
- \left (\bar \nabla\times{\bf A} \right )^2
\right )
dv \, ,
\label{EQq4.11}
\end{equation}
and a Lagrangian density,
\begin{equation}
{\cal L}=\frac{1}{2} \left (
\left (
\frac{\partial{\bf A}}{\partial \bar x^0}
\right )^2
- \left (\bar \nabla\times{\bf A} \right )^2
\right ) \, ,
\label{EQq4.12}
\end{equation}
in which the kinetic and potential terms are
explicitly quadratic corresponding to a conservative system with real
eigenvalues.
\par
In Lagrangian field theory, the Lagrangian density is not unique. 
In order to determine whether a given Lagrangian density is viable,
we must derive the Lagrange equations of motion and determine whether
the results agree with physical reality.
Specifically, it should not be asserted that the hypothesis,
Eq.~(\ref{EQq4.12}), of our Lagrangian field theory is a priori wrong
based on assumptions about the consequences before the theory is
actually developed.
Hundreds of years ago, it was wrong to assert that the hypothesis that
parallel lines meet is manifestly false.
Fortunately, non-Euclidian geometry managed to outlive its critics
although several generations of its proponents expired before it was
generally accepted.
More recently, Einstein's theory of special relativity is a purely
inductive theory that survived critics that advocated for `obvious'
or `well-established' absolute simultaneity \cite{BIEhrenfestQuote}.
\par
We take Eq.~(\ref{EQq4.12}) as our hypothesis and apply 
field theory to systematically derive equations of motion for 
macroscopic fields in an arbitrarily large simple linear
magneto-dielectric medium.
We develop a cohesive physical theory of field theory-based
electrodynamics, spacetime, relativity, tensor theory, and conservation
laws for a region of space in which the speed of light is $c/n$, rather
than $c$.
The new theory is demonstrated to be in agreement with the physical
world as is required.
\par
In this work, the index convention for Greek letters is that they belong
to $\{0,1,2,3\}$ and lower case Roman indices from the middle of the 
alphabet are in $\{1,2,3\}$. 
Cartesian coordinates $(x^1,x^2,x^3)$ correspond to $(x,y,z)$ as usual.
The Einstein summation convention in which repeated indices on the
same side of the equal sign are summed over is employed.
\par
\section{Lagrangian Equations of Motion}
\par
The Lagrange equations for electromagnetic fields in the vacuum
are \cite{BIGold,BICohen}
\begin{equation}
\frac{d}{d t}\frac{\partial{\cal L}}
{\partial (\partial A_j /\partial t)}
+\sum_i\frac{\partial}{\partial x_i}
\frac{\partial{\cal L}}{\partial (\partial A_j /\partial x_i)}
=\frac{\partial {\cal L}}{\partial A_j} \, .
\label{EQq5.01}
\end{equation}
We multiply and divide the first term of Eq.~(\ref{EQq5.01}) by
$c/n_e$ and the second term by $n_m$.
Using the re-parameterized temporal and spatial coordinates,
Eqs.~(\ref{EQq4.08}) and (\ref{EQq4.09}), the
preceding equation corresponds to
\begin{equation}
\frac{d}{d \bar x^0}\frac{\partial{\cal L}}
{\partial (\partial A_j /\partial \bar x^0)}
+\sum_i\frac{\partial}{\partial \bar x^i}
\frac{\partial{\cal L}}{\partial (\partial A_j /\partial \bar x^i)}
=\frac{\partial {\cal L}}{\partial A_j}
\label{EQq5.02}
\end{equation}
that we take as our hypothesis for simple linear magneto-dielectric
materials.
There is a different coordinate system
$\{\bar x^0,\bar x,\bar y,\bar z\}=
\{\bar x^0,\bar x^1,\bar x^2,\bar x^3\}$,
and a different set of Lagrange equations for each isotropic
homogeneous simple linear magneto-dielectric material.
A different non-Minkowski continuous spacetime
${\cal S}_c(\bar x^0,\bar x^1,\bar x^2,\bar x^3)$
is associated with each different coordinate system, see Sec. VII.
The vacuum Lagrange equations, Eqs.~(\ref{EQq5.01}), are a special case
of Eqs.~(\ref{EQq5.02}) for Minkowski spacetime.
\par
Substituting the density, Eq.~(\ref{EQq4.12}), into 
the Lagrange-like equations, Eq.~(\ref{EQq5.02}), we can
evaluate individual pieces as 
\begin{subequations}
\begin{equation}
\frac{\partial{\cal L}}
{\partial (\partial A_{j}/\partial \bar x^0)}
=\frac{\partial A_j}{\partial \bar x^0}
\label{EQq5.03a}
\end{equation}
\begin{equation}
\frac{\partial \cal L}{\partial A_j}=0
% \frac{n_eJ_j}{c}
\label{EQq5.03b}
\end{equation}
\begin{equation}
\sum_i\frac{\partial}{\partial \bar x^i}
\frac{\partial{\cal L}}{(\partial_{i} A_{j}/\partial \bar x^i)}
=[\bar\nabla\times(\bar\nabla\times {\bf A})]_j \, .
\label{EQq5.03c}
\end{equation}
\label{EQq5.03}
\end{subequations}
Substituting the individual pre-evaluated terms, Eqs.~(\ref{EQq5.03}),
back into Eq.~(\ref{EQq5.02}), the equations of motion for the
electromagnetic field in a simple linear magneto-dielectric medium
are the three orthogonal components of the vector wave equation
\begin{equation}
\bar \nabla\times(\bar \nabla\times {\bf A})
+\frac{\partial}{\partial\bar x^0}
\frac{\partial{\bf A}}{\partial\bar x^0}
=0 \;.
\label{EQq5.04}
\end{equation}
The second-order equation, Eq.~(\ref{EQq5.04}), can be written as a
set of first-order differential equations.
Writing portions of the wave equation as
\begin{subequations}
\begin{equation}
{\bf \Pi}=
\frac{\partial{\bf A}}{\partial \bar x^0}
\label{EQq5.05a}
\end{equation}
\begin{equation}
{\bf \bbeta}=\bar \nabla\times{\bf A} 
\label{EQq5.05b}
\end{equation}
\label{EQq5.05}
\end{subequations}
introduces macroscopic fields ${\bf \Pi}$ and ${\bf \bbeta}$.
The macroscopic field variable ${\bf \Pi}$, Eq.~(\ref{EQq5.05a}), is
the canonical momentum field \cite{BIGold} whose components were
derived as Eq.~(\ref{EQq5.03a}).
\par
We substitute the canonical momentum field ${\bf \Pi}$,
Eq.~(\ref{EQq5.05a}), and the magnetic field ${\bf \bbeta}$,
Eq.~(\ref{EQq5.05b}), into the wave equation, Eq.~(\ref{EQq5.04}), to
obtain
\begin{equation}
\bar\nabla\times{\bf \bbeta}
+\frac{\partial {\bf \Pi}}{\partial \bar x^0}
=0 \,,
%=\frac{\bar \nabla n_m}{n_m} \times {\bf \bbeta} \,,
\label{EQq5.06}
\end{equation}
which is similar in form to the Maxwell--Amp\`ere law in the
Maxwell--Minkowski representation of continuum electrodynamics,
but in a non-Minkowski `material' spacetime.
Applying the `material' divergence operator ($\bar\nabla\cdot$) to
Eq.~(\ref{EQq5.05b}), we obtain
\begin{equation}
\bar\nabla\cdot{\bf \bbeta}
= 0 \, .
%= -\frac{\bar\nabla n_m}{n_m} \cdot {\bf \bbeta} \, .
\label{EQq5.07}
\end{equation}
Applying the `material' curl operator ($\bar\nabla\times$) to
Eq.~(\ref{EQq5.05a}) produces a version of Faraday's Law,
\begin{equation}
\bar\nabla\times{\bf \Pi}
-\frac{\partial{\bf \bbeta}}{\partial \bar x^0}
= 0\, .
%= \frac{\bar\nabla n_e}{n_e}\times{\bf \Pi} \, .
\label{EQq5.08}
\end{equation}
Finally,
\begin{equation}
\bar\nabla\cdot{\bf \Pi}
= 0
%= -\frac{\bar\nabla n_e}{n_e}\cdot {\bf \Pi} 
\label{EQq5.09}
\end{equation}
is a modified version of Gauss's law that is obtained by integrating 
the material divergence of Eq.~(\ref{EQq5.06}) with respect to the new
timelike coordinate $\bar x^0$.
This completes the set of first-order equations of motion,
Eqs.~(\ref{EQq5.06})--(\ref{EQq5.09}), for macroscopic fields in an
arbitrarily large, isotropic, homogeneous, simple linear
magneto-dielectric medium.
\par
Grouping the field equations, Eqs.~(\ref{EQq5.06})--(\ref{EQq5.09}),
for clarity and convenience, we have
\begin{subequations}
\begin{equation}
\bar\nabla\times{\bf \bbeta}
+\frac{\partial {\bf \Pi}}{\partial \bar x^0}
%=\frac{\bar \nabla n_m}{n_m} \times {\bf \bbeta}
=0 
\label{EQq5.10a}
\end{equation}
\begin{equation}
\bar\nabla\cdot{\bf \bbeta}
%-\frac{\bar\nabla n_m}{n_m} \cdot {\bf \bbeta}
=0
\label{EQq5.10b}
\end{equation}
\begin{equation}
\bar\nabla\times{\bf \Pi}
-\frac{\partial{\bf \bbeta}}{\partial \bar x^0}
%= \frac{\bar\nabla n_e}{n_e}\times{\bf \Pi}
=0
\label{EQq5.10c}
\end{equation}
\begin{equation}
\bar\nabla\cdot{\bf \Pi}=0
%-\frac{\bar\nabla n}{n}\cdot {\bf \Pi}
\label{EQq5.10d}
\end{equation}
\label{EQq5.10}
\end{subequations}
as the equations of motion for macroscopic electromagnetic fields
in an arbitrarily large isotropic, homogeneous, simple linear
magneto-dielectric medium.
\par
Equations~(\ref{EQq5.10}) appear to violate Einstein's relativity,
except Einstein's relativity was derived for events in the vacuum
of Minkowski spacetime.
On the other hand, Eqs.~(\ref{EQq5.10}) are fully consistent with
Rosen's dielectric special relativity \cite{BIRosen} 
for an observer in the non-Minkowski continuous `material' spacetime
that is associated with a linear medium \cite{BIRosen,BIAJP}, see
Sec.~VI.
The version of dielectric special relativity that was derived
by Laue \cite{BILaue} using the relativistic velocity sum rule
applies to an observer in a Laboratory Frame of Reference that
is in the vacuum (or tenuous terrestrial atmosphere) that
surrounds the dielectric \cite{BIAJP}.
\par
Equations~(\ref{EQq5.10}) will also apply to a piecewise-homogeneous
material with Fresnel boundary conditions \cite{BINewFres}.
The vacuum Maxwell equations correspond to a special case of
Eqs.~(\ref{EQq5.10}) in Minkowski spacetime.
\par
The continuum limit is a theoretical abstraction in which the linear
medium is isotropic, homogeneous, and continuous at all length scales
from the very outset.
This is represented in field theory by defining both the Lagrange
equations, Eq.~(\ref{EQq5.02}), and the Lagrangian density,
Eq.~(\ref{EQq4.12}), for the non-Minkowski continuous
`material spacetime' ${\cal S}_c(\bar x^0,\bar x^1,\bar x^2,\bar x^3)$,
see Sec. VII.
Because each linear material is associated with its own material 
spacetime, there is no need for independent material parameters
like the permittivity and permeability (except as boundary conditions
to identify and relate different spacetimes).
Then, the Maxwell--Minkowski equations cannot be derived 
as an identity of Eqs.~(\ref{EQq5.10}) despite having derived
the similar appearing Eqs.~(\ref{EQq2.28})
as an identity of the Maxwell--Minkowski equations in
Refs.~\cite{BIIdentity,BIJMP}.
\par
Fundamental physical processes are derived and defined for the vacuum.
The microscopic Maxwell equations are fundamental laws of
electrodynamics in the vacuum.
Maxwellian continuum electrodynamics, which is obtained by adding a
phenomenological material response as a perturbation of the vacuum
Maxwell equations, has been known to be inconsistent with
conservation laws for over a century and was proven to be
manifestly false in Sec. I.
Therefore, no formula, theorem, or other result of Maxwellian continuum
electrodynamics can be used to disprove the new field equations,
Eqs.~(\ref{EQq5.10}), that are derived in an isotropic, homogeneous,
flat, four-dimensional, non-Minkowski, continuous spacetime 
${\sf S}_c(\bar x^0,\bar x^1,\bar x^2,\bar x^3)$.
Other fundamental physical processes like relativity and conservation
are likewise rooted in the vacuum and phenomenologically transported
into the continuous linear medium with a view to consistency with 
Maxwellian continuum electrodynamics.
Consequently, these processes cannot be used to disprove the new
field equations, Eqs.~(\ref{EQq5.10}), either.
The process of recasting these fundamental principles for a
medium that is continuous at all length scales from the outset
has begun and we can report success in demonstrating that
Eqs.~(\ref{EQq5.10}) are consistent with the Fresnel
relations \cite{BINewFres} and special relativity in a
dielectric \cite{BIAJP}.
\par
\section{Special Relativity in a Magneto-Dielectric Medium}
\par
In adapting Einstein's special theory of relativity to a dielectric
medium, Laue \cite{BILaue} applied the relativistic velocity sum rule
to a dielectric material moving uniformly in the vacuum-based local
Laboratory Frame of Reference.
Some four decades later, Rosen \cite{BIRosen} considered a continuous
dielectric medium that is sufficiently large that the vacuum is
inaccessible from the interior (in the time it takes for an experiment
to be performed) and derived a second form of dielectric special
relativity by a phenomenological replacement of the speed of light
by $c/n$.
The phenomenological Rosen theory and its consequences are mostly
ignored in the scientific literature and there is little or no
discussion about the incompatibility of the two contradictory
theories of relativity in a dielectric \cite{BIAJP}.
\par
In Ref. \cite{BIAJP}, the current author used two inertial
reference frames in uniform motion to prove that the two forms of
dielectric special relativity are incommensurate and that both are
correct, but are correct in different physical contexts.
Placing the common origin of the inertial reference frames on the
interface between a semi-infinite medium and the vacuum and restricting
the direction of relative motion to the interface \cite{BIAJP},
it was found that the Laue \cite{BILaue} theory, with `vacuum'
Lorentz factor
\begin{equation}
\gamma_v=\sqrt{\frac{1}{1-v^2/c^2}} \, ,
\label{EQq6.01}
\end{equation}
relates electrodynamics inside the material to a vacuum-based local
Laboratory Frame of Reference outside of the material.
In contrast, the Rosen \cite{BIRosen} theory, with `material' Lorentz
factor,
\begin{equation}
\gamma_d=\sqrt{\frac{1}{1-\varepsilon v^2/c^2}} \, ,
\label{EQq6.02}
\end{equation}
applies if both inertial reference frames are within an arbitrarily
large isotropic, homogeneous, simple linear dielectric medium.
In this section, we extend the theory of dielectric special
relativity \cite{BIAJP} to include simple linear magneto-dielectric
 media.
\par
%\begin{figure}
%\includegraphics[scale=0.72]{New2Crop.pdf}
%\caption{Coordinate frame $\bar S$ in the dielectric.}
%\label{fig1}
%\end{figure}
We consider two inertial frames of reference,
$\bar S(\bar x,\bar y,\bar z)$ and
$\bar S^{\prime} (\bar x^{\prime},\bar y^{\prime},\bar z^{\prime})$,
in a standard configuration.
The origins of the reference frames are located inside an
arbitrarily large isotropic homogeneous simple linear
magneto-dielectric medium.
The origins of the two systems coincide at time $t_0=0$ and
all clocks are synchronized.
At time $t_c=t_c^{\prime}=0$, a light pulse is emitted from the common
origin along the positive $\bar y$ and $\bar y^{\prime}-$axes.
In the $\bar S$ frame of reference, Fig.~1, the pulse is reflected by
a mirror in the medium at $\bar y=D_c$ and returns to the origin at
time $\Delta t_c=2D_c/c_c$, where $c_c$ is the speed of light in the
medium.
The trajectory of the light pulse in the $\bar S^{\prime}$ frame of 
reference is shown in Fig.~2.
The translation of the $\bar S^{\prime}$ frame is transverse to the
$\bar y$-axis so the distance from the mirror at $m_c^{\prime}$ to the
$\bar x^{\prime}$-axis is $D_c$, the same as the distance from the
mirror at $m_c$ to the $\bar x$-axis.
Viewed from the $\bar S^{\prime}$ frame, the light pulse is emitted from
the point $o$ at time $t_c^{\prime}=0$, is reflected from the mirror at
point $m_c^{\prime}$, and is detected at the point $d_c^{\prime}$ at
time $t_c^{\prime} =\Delta t_c^{\prime}$.
During that time, the point of emission/detection has moved a
distance $v_c\Delta t_c^{\prime}$.
\par
By the Pythagorean theorem, we have
\begin{equation}
(c_c^{\prime}\Delta t_c^{\prime})^2
=(c_c\Delta t_c)^2+(v_c\Delta t_c^{\prime})^2 \, ,
\label{EQq6.03}
\end{equation}
where we have used reflection symmetry about the midpoint.
We write the previous equation as \cite{BIAJP}
\begin{equation}
\Delta t_c^{\prime}=
\frac{\Delta t_c}{\sqrt{{c_c^{\prime}}^2/c_c^2-v_c^2/c_c^2}}
\label{EQq6.04}
\end{equation}
and define the `material' Lorentz factor $\gamma_c$ by
\begin{equation}
\Delta t_c^{\prime}= \gamma_c \Delta t_c
\label{EQq6.05}
\end{equation}
such that
\begin{equation}
\gamma_c= \frac {1}{\sqrt{{c_c^{\prime}}^2/c_c^2-v_c^2/c_c^2}} \,.
\label{EQq6.06}
\end{equation}
\par
In the Laue model, the speed of light in the dielectric depends on the
velocity of the block in the Laboratory Frame of Reference.
Here, the isotropy of an arbitrarily large homogeneous continuous
dielectric medium at rest in the local frame of reference leads us to
postulate that light travels at a uniform speed $c_c$ in the simple
linear magneto-dielectric medium, basically the same reasoning that
led to the Einstein postulate.
Substituting
\begin{equation}
c_c^{\prime}=c_c
\label{EQq6.07}
\end{equation}
into Eq.~(\ref{EQq6.06}), one obtains
\begin{equation}
\gamma_c= \frac {1}{\sqrt{1- v_c^2/c_c^2}} \,.
\label{EQq6.08}
\end{equation}
Using the definitions of the renormalized coordinates, we have
$v_c=n_mv$ and $c_c=c/n_e$.
Then,
\begin{equation}
\gamma_c=\frac{1}{\sqrt{1- n_e^2n_m^2v^2/c^2}}
=\frac{1}{\sqrt{1- n^2v^2/c^2}}
=\frac{1}{\sqrt{1- v^2/\bar c^2}}
\label{EQq6.09}
\end{equation}
for a simple linear magneto-dielectric medium ${\bar c}=c/n$.
There is a different material Lorentz factor for each isotropic
homogeneous simple linear magneto-dielectric medium.
The vacuum Lorentz factor, Eq.~(\ref{EQq6.01}), is a special case of
Eq.~(\ref{EQq6.09}) for $n_e=n_m=1$.
The Rosen dielectric Lorentz factor, Eq.~(\ref{EQq6.02}), is a special
case of Eq.~(\ref{EQq6.09}) for $n_m=1$.
\par
\section{Spacetime Setting}
\par
If a light pulse is emitted from the origin at a time $t=0$ into the
empty vacuum of free space then spherical wavefronts are defined by 
\begin{equation}
x^2+y^2+z^2=(x^0)^2
\label{EQq7.01}
\end{equation}
in a flat, four-dimensional, vacuum Minkowski spacetime
${\sf S}_M(x^0=ct,x,y,z)$.
Equation (\ref{EQq7.01}) underlies classical electrodynamics and its
relationship to Einstein's special relativity.
\par
Consider a quasimonochromatic light pulse that is emitted from the
origin $(\bar x=0,\bar y=0,\bar z=0)$ at `time' $\bar x^0=0$ into an
isotropic, homogeneous, simple linear magneto-dielectric medium,
instead of the vacuum.
In this medium, spherical wavefronts are defined by
\begin{equation}
\bar x^2+\bar y^2+\bar z^2=(\bar x^0)^2
\label{EQq7.02}
\end{equation}
in an isotropic, homogeneous, flat, four-dimensional, non-Minkowski
continuous material
spacetime ${\sf S}_c(\bar x^0,\bar x,\bar y,\bar z)$.
There will be a different material spacetime that is associated with
each set of refractive indices, $n_e$ and $n_m$.
\par
The four-dimensional `material' light cone 
\begin{equation}
\bar x^2+\bar y^2+\bar z^2-(\bar x^0)^2=0 
\label{EQq7.03}
\end{equation}
is embedded in the flat four-dimensional non-Minkowski `material'
spacetime ${\sf S}_c(\bar x^0,\bar x,\bar y,\bar z)$ that is
associated with a linear, isotropic, homogeneous, simple linear
magneto-dielectric medium.
The basis functions,
$\exp(-i(\bar\omega/c)(\bar x^0-{\bf \hat k}\cdot{\bf \bar r}))+c.c.$,
where ${\bf\bar r}=(\bar x,\bar y,\bar z)$,
define the null surface, $\bar x^0={\bf \hat k}\cdot{\bf \bar r}$.
Here, $\bar \omega$ can be associated with $n_e \omega$.
Fig.~3 is a depiction of the intersection of the `material' light cone
with the $\bar x^0-\bar x$ plane in the flat material spacetime showing
the null $\bar x^0=\bar x$.
%\begin{figure}
%\includegraphics[scale=0.22]{FigXXX.pdf}
%\label{fieldfig3}
%\end{figure}
There will be a different material spacetime for each pair of
material constants, $n_e$ and $n_m$, but the half-opening angle of the
material light cone will always be $\alpha=\pi/4$ in that spacetime.
The unit slope of the null in the $\bar x^0-\bar x$ plane of the
non-Minkowski material spacetime is related to the coordinate speed of
light in an isotropic, homogeneous, simple linear medium by
\begin{equation}
\frac{d x}{d t}=
\frac{d\bar x}{d\bar x^0}
\frac{d\bar x^0}{dt}
\frac{dx}{d\bar x} =
1\cdot\frac{c}{n_e}\cdot\frac{1}{n_m} =\frac{c}{n_en_m}
=\frac{c}{n} =\bar c \, .
\label{EQq7.04}
\end{equation}
This equation shows that the effective speed of light in a
linear magneto-dielectric medium is attributable to two different
effects:
{\it i}) the renormalization of the timelike coordinate by $n_e^{-1}$
and {\it ii}) the renormalization of the spatial coordinates by $n_m$.
\par
The usual characterization of events inside the vacuum light cone as
timelike and events outside the vacuum light cone as spacelike also 
applies to the `material' spacetime with events inside the renormalized
`material' light cone being timeline and events outside are spacelike.
\v{C}erenkov radiation, spontaneous emission, and mass-bearing particle
dynamics in a simple linear medium will have to be treated carefully,
if at all, because an atom or a charged particle must displace some
of the linear medium that is effectively continuous at all length
scales.
\par
\begin{figure}
\includegraphics[scale=0.70]{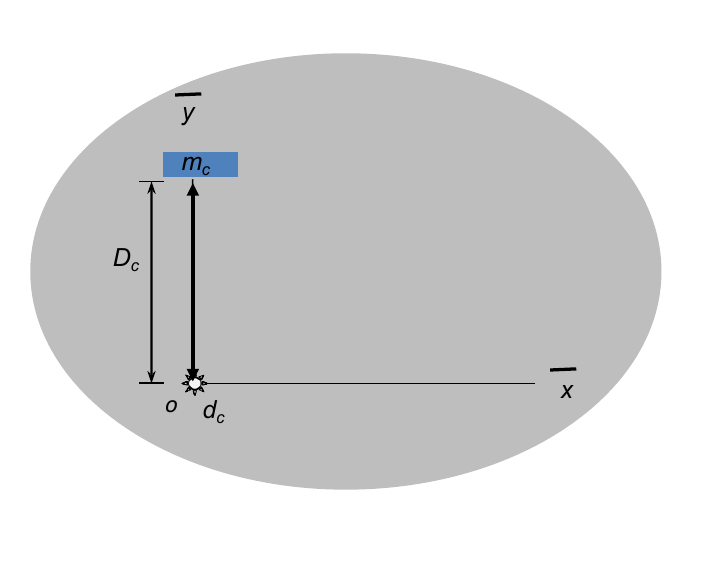}
\caption{Path of light in the unprimed coordinate frame.}
%\label{fig3}
\end{figure}
\begin{figure}
\includegraphics[scale=0.70]{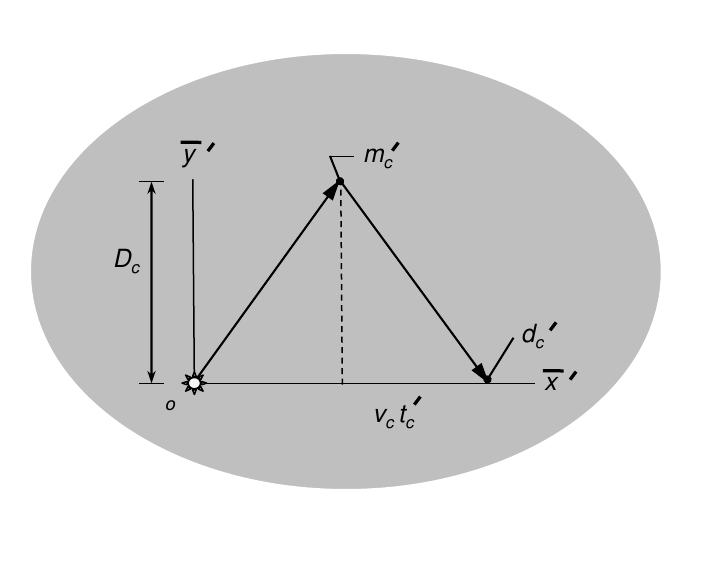}
\caption{Path of light in the primed coordinate frame.}
%\label{fig3}
\end{figure}
\par
\begin{figure}
\includegraphics[scale=0.60]{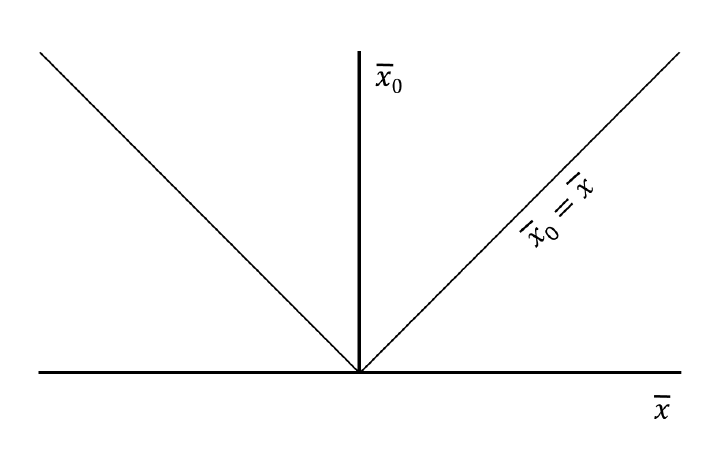}
\caption{Null cone for light depicted in the $\bar x^0-\bar x$ plane
of a flat, non-Minkowski material spacetime that corresponds to a
linear magneto-dielectric medium.}
%\label{fig3}
\end{figure}
\par
\section{Lorentz transformation properties}
\par
Maxwellian continuum electrodynamics is based on averaging the
interaction of microscopic fields with tiny bits of polarizable 
and magnetizable matter embedded in the vacuum.
In the vacuum, Lorentz transformations of the coordinates follow from 
the invariance of
\begin{equation}
s^2=(x^0)^2-x^2-y^2-z^2 \, .
\label{EQq8.01}
\end{equation}
Then it makes sense to apply Lorentz invariance to the microscopic 
theory before the interactions are averaged in the continuum 
approximation and to apply Lorentz invariance to `real' 
magneto-dielectric materials that are comprised almost entirely 
of empty space.
Except, a macroscopic linear medium is defined as being continuous
at all length scales \textit{from the very outset} because the
performing of such a macroscopic average from microscopic particles,
fields, and interactions requires assumptions, approximations,
and limiting behavior that become dubious in their complexity.
Consequently, the assumption of Lorentz invariance for a macroscopic
linear medium appears to be well-founded, based on the microscopic
model, but
it is manifestly not well-founded for macroscopic fields in continuous
linear media.
\par
By now, we should know that Lorentz invariance is not a symmetry of
light in a macroscopic linear medium \cite{BIFinn} and we cannot
re-discretize or un-average the physics in order to impose Lorentz
invariance.
Instead, the invariance of 
\begin{equation}
(\bar s )^2=(\bar x^0)^2-(\bar x^1)^2-(\bar x^2)^2-(\bar x^3)^2
\label{EQq8.02}
\end{equation}
for monochromatic light in an arbitrarily large,
isotropic, homogeneous simple linear medium imposes the conditions for
linear medium-specific Lorentz-like transformations of the coordinates
$(\bar x^0,\bar x^1,\bar x^2,\bar x^3)$ of a flat four-dimensional
non-Minkowski material 
spacetime ${\sf S}_c(\bar x^0,\bar x,\bar y,\bar z)$.
The special Lorentz-like transformations (with ${\bf v}$ parallel to
the $\bar x$-axis) take the form \begin{subequations}
\begin{equation}
\bar x^{0^{\prime}}=\gamma_c\left (\bar x^0-(n_en_mv/c)\bar x^1\right )
=\gamma_c\left ( \bar x^0-(nv/c)\bar x^1 \right )
\label{EQq8.03a}
\end{equation}
\begin{equation}
\bar x^{1^{\prime}}=\gamma_c\left (\bar x^1-(n_en_mv/c)\bar x^0\right )
=\gamma_c\left ( \bar x^1-(nv/c)\bar x^0 \right )
\label{EQq8.03b}
\end{equation}
\begin{equation}
\bar x^{2^{\prime}}=\bar x^2
\label{EQq8.03c}
\end{equation}
\begin{equation}
\bar x^{3^{\prime}}=\bar x^3
\label{EQq8.03d}
\end{equation}
\label{EQq8.03}
\end{subequations}
in a simple linear medium.
The index-dependent Lorentz-like transformation confirms the
observations of Ravndal \cite{BIFinn} that the invariance properties
of a linear medium differ from Lorentz invariance of the vacuum.
\par
\section{Tensor Formulation}
\par
With the advent of special relativity, the 3-vector formulation of
electrodynamics became archaic.
The tensor formalism allows the development of electrodynamics in the
context of general properties of physical laws on Minkowski spacetime 
in a form that is manifestly invariant under Lorentz transformations.
``Increasing the sophistication of the notation simplifies the
appearance of the governing equations, revealing hidden symmetries
and deeper meaning in the equations of
electromagnetism'' \cite{BIWarnickR}.
\par
The fundamental laws of physics are formulated in the vacuum and
Minkowski spacetime is empty.
That is not considered to be a problem for continuous media
because `real' matter is mostly empty space.
Except, Feynman's \cite{BIFeynman} pedagogy makes it clear
that there is a place in physics for macroscopic descriptions of
fields and matter.
Physical theory that is formulated for enumerated localized
microscopic particles and microscopic fields interacting in a
vacuum and defined on an empty Minkowski spacetime is always
correct, but it is manifestly unjustified for macroscopic fields
in an effective medium that is isotropic, homogeneous, and
continuous at all length scales from the very outset.
\par
In this section, we will develop an expressly macroscopic tensor
formulation of continuum electrodynamics based on the macroscopic field
equations, Eqs.~(\ref{EQq5.10}).
\par
It is straightforward to use algebra and calculus in order to construct
the energy--momentum tensor as a theorem of the macroscopic field
equations.
The derivation of the energy continuity equation follows the same 
procedure that was used to derive Poynting's theorem,
Eq.~(\ref{EQq2.09}), as an identity of the Maxwell--Minkowski equations.
We combine Eqs.~(\ref{EQq5.10}) 
to derive a theorem of macroscopic continuum electrodynamics in the
form the scalar energy continuity equation
\begin{equation}
\frac{\partial{u_c}}{\partial \bar x^0} 
+\bar\nabla\cdot{\bf s}_c =0 \, ,
\label{EQq9.01}
\end{equation}
in the continuous material spacetime
where
\begin{equation}
u_c=\frac{1}{2} ( {\bf \Pi}^2+{\bbeta}^2) 
\label{EQq9.02}
\end{equation}
is the continuous energy density and
\begin{equation}
{\bf s}_c=c{\bbeta}\times{\bf \Pi}
\label{EQq9.03}
\end{equation}
is the continuous `material' energy-flux vector.
Similarly we combine Eqs.~(\ref{EQq5.10}) to form the vector momentum
continuity theorem
\begin{equation}
\frac{\partial g^i_c}{\partial \bar x^0}
+\sum_j\frac{\partial}{\partial \bar x^j}{{\sf W}_c}^{ij}=0
\label{EQq9.04}
\end{equation}
in component form, where
\begin{equation}
{\bf g}_c=\frac{{\bbeta}\times{\bf \Pi}}{c}
\label{EQq9.05}
\end{equation}
denotes the `material' momentum density and
\begin{equation}
{{\sf W}_c}^{ij}=-\Pi^i\Pi^j-\beta^i\beta^j+
\frac{1}{2} ({\bf \Pi}^2+{\bbeta}^2)\delta^{ij}
\label{EQq9.06}
\end{equation}
are the elements of a rank 3 matrix.
Readers that are not familiar with the procedure used to derive
Eq.~(\ref{EQq9.04}) can consult Ref.~\cite{BIBoydMil},
Sec.~6.8 of Ref.~\cite{BIJackson}, or similar reference.
\par
The scalar energy continuity equation, Eq.~(\ref{EQq9.01}),
and the scalar components of the vector momentum continuity
equation, Eqs.~(\ref{EQq9.04}), can be
written, row-wise, as a single differential equation \cite{BIOptCommun}
\begin{equation}
\bar \partial_{\beta} {\sf T}_c^{\alpha\beta} =0
\label{EQq9.07}
\end{equation}
as a matter of linear algebra, where
\begin{equation}
\bar \partial_{\beta} =\left (\frac{\partial}{\partial \bar x^0},
\bar\nabla\right ) 
\label{EQq9.08}
\end{equation}
is the `material' four-divergence operator.
The `tensor' differential equation, Eq.~(\ref{EQq9.07}), is a theorem
of the macroscopic field equations, Eqs.~(\ref{EQq5.10}),
and the diagonally symmetric matrix ${\sf T}_c^{\alpha\beta}$ is
\begin{equation}
{\sf T}_c^{\alpha\beta}
\! = \!
\left [
\begin{matrix}
{u}    \!&{s}_c^1/c        \!&{s}_c^2/c           \!&{s}_c^3/c
\cr
c{g}_c^1   &{\sf W}_c^{11}   &{\sf W}_c^{12}      &{\sf W}_c^{13}
\cr
c{g}_c^2   &{\sf W}_c^{21}   &{\sf W}_c^{22}      &{\sf W}_c^{23}
\cr
c{g}_c^3   &{\sf W}_c^{31}   &{\sf W}_c^{32}      &{\sf W}_c^{33}
\cr
\end{matrix}
\right ] \, ,
\label{EQq9.09}
\end{equation}
by construction.
Obviously, the intent is to identify ${\sf W}_c^{ij}$ as the continuum
electrodynamic stress tensor, to identify ${\sf T}_c^{\alpha\beta}$
as the continuum electrodynamic energy--momentum tensor, and to
identify the differential equation, Eq.~(\ref{EQq9.07}),
with the local electromagnetic conservation law.
\par
We can re-formulate most of the other features of tensor
electrodynamics, c.f., Sec.~11 of Ref.~\cite{BIJackson}.
We construct the field strength tensor
\begin{equation}
{\sf F}_c^{\alpha\beta}
\! = \!
\left [
\begin{matrix}
0            \!& {\bf \Pi}_x  \!& {\bf \Pi}_y      \!& {\bf \Pi}_z
\cr
-{\bf \Pi}_x   &0               &-{\bbeta}_z         &{\bbeta}_y
\cr
-{\bf \Pi}_y   &{\bbeta}_z      &0                   &-{\bbeta}_x
\cr
-{\bf \Pi}_z   &-{\bbeta}_y     &{\bbeta}_x          &0                
\cr
\end{matrix}
\right ]
\label{EQq9.10}
\end{equation}
and the dual field-strength tensor
\begin{equation}
{\cal F}_c^{\alpha\beta}
\! = \!
\left [
\begin{matrix}
0            \!&-{\bbeta}_x  \!&-{\bbeta}_y      \!&-{\bbeta}_z
\cr
{\bbeta}_x   &0               &-{\bf\Pi}_z         &{\bf\Pi}_y
\cr
{\bbeta}_y   &{\bf\Pi}_z      &0                   &-{\bf\Pi}_x
\cr
{\bbeta}_z   &-{\bf\Pi}_y     &{\bf\Pi}_x          &0 
\cr
\end{matrix}
\right ] 
\label{EQq9.11}
\end{equation}
such that
\begin{subequations}
\begin{equation}
\bar\partial_{\alpha} {\sf F}_c^{\alpha\beta}=0
\label{EQq9.12a}
\end{equation}
\begin{equation}
\bar\partial_{\alpha} {\cal F}_c^{\alpha\beta}=0
\label{EQq9.12b}
\end{equation}
\label{EQq9.12}
\end{subequations}
constitute an identity of the macroscopic field equations,
Eqs.~(\ref{EQq5.10}).
The energy--momentum tensor, Eq.~(\ref{EQq9.09}),
\begin{equation}
{\sf T}_c^{\alpha\beta}=
{\sf F}^{\alpha\mu}{\sf F}^{\beta}_{\hskip 0.04in \mu }
+\frac{1}{4}g^{\alpha\beta}{\sf F}_{\mu\sigma}{\sf F}^{\mu\sigma}
\label{EQq9.13}
\end{equation}
and the Lagrangian density, Eq.~(\ref{EQq4.12}),
\begin{equation}
{\cal L}_c=\frac{1}{2}\left ( {\bf \Pi}^2-{\bbeta}^2\right )
= -\frac{1}{4} {\sf F}_{\alpha\beta}{\sf F}^{\alpha\beta}
\label{EQq9.14}
\end{equation}
are easily demonstrated by substitution of the field strength
tensor, Eq.~(\ref{EQq9.10}).
Combining the definitions Eq.~(\ref{EQq9.05}) and
Eq.~(\ref{EQq9.03}) we obtain 
\begin{equation}
{\bf s}=c^2{\bf g} \; .
\label{EQq9.15}
\end{equation}
The trace of the energy--momentum tensor vanishes
\begin{equation}
g_{\alpha\alpha} T_c^{\alpha\alpha}=0
\label{EQq9.16}
\end{equation}
corresponding to a continuous flow of massless particles.
Substituting ${\bf s}=c^2{\bf g}$ into the energy--momentum tensor,
Eq.~(\ref{EQq9.09}), we have the symmetry property
\begin{equation}
{\sf T}_c^{\alpha\beta} = {\sf T}_c^{\beta\alpha} \, .
\label{EQq9.17}
\end{equation}
Using the symmetry property, Eq.~(\ref{EQq9.17}), of the
energy--momentum tensor, we obtain
\begin{equation}
\bar \partial_{\alpha} {\sf T}_c^{\alpha\beta} =0
\label{EQq9.18}
\end{equation}
from Eq.~(\ref{EQq9.07}).
\par
Returning to Maxwellian continuum electrodynamics for a moment,
the well-known Minkowski energy--momentum tensor can be constructed
by linear algebra from macroscopic energy and momentum continuity
equations that are theorems of the Maxwell--Minkowski equations,
but the Minkowski energy--momentum tensor is not symmetric.
The Faraday law, Eq.~(\ref{EQq2.03b}), in a linear medium 
\begin{equation}
\nabla\times{\bf E} -\frac{\partial{\bf B}}{\partial (ct)} =0
\label{EQq9.19}
\end{equation}
has the same timelike coordinate as the Faraday law in the vacuum
Minkowski spacetime, $S_v(x^0=ct,z,y,z)$.
In contrast, the Maxwell--Amp\`ere law, Eq.~(\ref{EQq2.03a}),
\begin{equation}
\nabla\times{\bf B} +\frac{\partial {\bf E}}{\partial (ct/n^2)} =0 
\label{EQq9.20}
\end{equation}
has a renormalized temporal coordinate that we would associate with
the non-Minkowski spacetime, $S_{?}(ct/n^2,z,y,z)$.
These equations cannot be combined self-consistently to form valid
energy and momentum continuity equations because they are based on
different coordinate systems and belong in different spacetimes.
The Minkowski energy-momentum tensor is not symmetric because
the macroscopic Maxwell--Minkowski equations are inconsistently
defined in a pathological spacetime.
In contrast, all of the new macroscopic field equations are in the
continuous `material' spacetime,
${\sf S}_c(\bar x^0,\bar x,\bar y,\bar z)$ that is consistent with
the electric and magnetic properties of the linear medium.
\par
\section{Experimental Confirmation}
\par
\subsection{The Balazs thought experiment}
\par
In 1953, Balazs \cite{BIBalazs} proposed a thought experiment to
resolve the Abraham--Minkowski controversy.
The thought experiment was based on the law of conservation of momentum
and a theorem that the center of mass-energy moves at a uniform
velocity \cite{BIBoyer}.
The application of this theorem indicates that microscopic constituents
of the material that carry mass also travel with the field.
\par
The relativistic total energy
\begin{equation}
E=\left ( {\bf p}\cdot{\bf p}c^2+m^2c^4 \right )^{1/2}
\label{EQq10.01}
\end{equation}
becomes the Einstein formula $E=mc^2$ for massive particles
in the limit ${\bf v}/c \rightarrow 0$.
For massless particles, like photons, Eq.~(\ref{EQq10.01}) becomes
\begin{equation}
{\bf p}=\frac{E}{c} {\bf \hat e}_k =
\frac{\hbar\omega_0}{c} {\bf \hat e}_k \, ,
\label{EQq10.02}
\end{equation}
where ${\bf \hat e}_k$ is a unit vector in the direction of motion.
Equation (\ref{EQq10.02}) defines the instantaneous momentum of a
photon traveling at speed $c$.
\par
Consider a photon that enters a material that is composed of electric
and magnetic dipoles embedded in the vacuum.
The photon travels at speed $c$ between scattering
events \cite{BIFeynman}, but due to scattering the effective speed
of the photon in the incident direction is $c/n$.
Due to the reduced effective speed of individual photons, an unimpeded,
inviscid, incoherent, incompressible flow of non-interacting photons in
the continuum limit travels at an average speed of $c/n$ and the
longitudinal extent of the field in the direction of the flow is
reduced by a factor of $n$.
Then the photon density, the energy density, and the momentum density
are increased by a factor $n$.
Integrating the quantities over the reduced longitudinal width of the
field, the energy and momentum are constant as the field leaves vacuum
and enters a dielectric medium without needing to assume any material
motion.
Due to the decreased velocity and increased density, 
the energy velocity of the ensemble is $c=n\cdot c/n$.
Then the energy velocity of a field into and through a linear medium
moves at a constant speed $c$ because of the enhanced photon density.
The mass-polariton (MP) model \cite{BIPart} of propagation of the
electromagnetic field in which the propagating field combines with
mass-bearing particles of a continuous dielectric is not supported.
Because the density of photons is larger due to the reduced average
velocity, the higher density of photons corresponds to an increase
in energy density and momentum density in the macroscopic field.
Then, the macroscopic field in a linear medium that corresponds to
the energy of a single photon occupies a smaller volume than the
volume occupied in the vacuum.
Likewise, a smaller volume of the field is associated with the
momentum of a single photon.
An additional issue with the photon description of light propagation
in a continuous dielectric is illustrated by the commingling of
macroscopic fields and the macroscopic refractive index with
microscopic photon momentum and momentum states in a description of
photon recoil momentum in a medium \cite{BICampbell}.
\par
As an electromagnetic field propagates from vacuum into a simple
linear medium, the `effective' velocities of photons in the field
are reduced creating an enhancement of the classical energy
density $u_c=({\bf \Pi}^2+{\bbeta}^2)/2$ and the classical momentum
density ${\bf g}_c={\bbeta}\times{\bf \Pi}/c$, compared to the vacuum.
For finite pulses in a dielectric, the enhanced energy density is 
offset by a narrowing of the pulse so that the electromagnetic energy 
\begin{equation}
U_{total}=\int_{\Sigma}
\frac{1}{2}\frac{{\bf\Pi}^2+{\bbeta}^2}{c}dv \, ,
\label{EQq10.03}
\end{equation}
is time independent for quasimonochromatic fields in the plane-wave
limit.
The electromagnetic energy is the total energy by virtue of being
constant in time.
Likewise, the electromagnetic momentum,
\begin{equation}
{\bf G}_{total}=\int_{\Sigma} \frac{{\bbeta}\times{\bf \Pi}}{c} dv \,,
\label{EQq10.04}
\end{equation}
is time independent and is the total momentum.
\par
Invoking the Einstein mass--energy equivalence, it is argued that
some microscopic constituents of the dielectric must be accelerated
and then decelerated by the field;
otherwise the theorem that the center of mass--energy moves at a
constant velocity is violated \cite{BIBarn}.
For a distribution of particles of mass $m_i$ and velocity ${\bf v}_i$,
the total momentum
\begin{equation}
{\bf P}_{total}= \sum_i m_i {\bf v}_i 
\label{EQq10.05}
\end{equation}
is the sum of the momentums of all the particles $i$ in the
distribution.
If the mass of each particle $m_i$ is constant, the statement that the
velocity of the center of mass
\begin{equation}
{\bf v}_{CM}= \frac{\sum_i m_i {\bf v}_i}{\sum_i m_i}
\label{EQq10.06}
\end{equation}
is constant is a statement of conservation of total momentum.
\par
Because of the enhanced momentum density of the field in a dielectric,
the differential of electromagnetic momentum
\begin{equation}
\delta {\bf p}=\frac{{\bbeta}\times{\bf \Pi}}{c} \delta v
\label{EQq10.07}
\end{equation}
that is contained in an element of volume $\delta v$ (a `particle'),
is a factor of $n$ greater than in the vacuum.
Then the mass--energy of each `particle' $m_i$ is not constant as
would be required for the center-of-mass theorem.
For a finite pulse, the narrower pulse width offsets the enhanced
momentum density allowing the macroscopic electromagnetic
momentum, like the macroscopic electromagnetic energy, to be
constant in time as the field enters, and exits, the simple linear
medium through the gradient-index antireflection coating.
Consequently, there is no need to hypothesize mass-polariton
quasiparticles \cite{BIPart} or any other material constituents of
the continuous linear medium to be in motion in order to preserve
the conservation of linear momentum.
Even though the velocity of light slows to $c/n$, the hypothetical
Minkowski pull-force is also disproved.
\par
\subsection{The Jones--Richards experiment}
\par
One of the enduring questions of the Abraham--Minkowski controversy is
why the Minkowski momentum is so often measured experimentally while
the Abraham form of momentum is so favored in theoretical work.
We now have the tools to answer that question.
The Minkowski momentum is not measured directly, but inferred from a
measured index dependence of the optical force on a mirror placed in a
dielectric fluid \cite{BIPfei,BIBarnLou,BIExp}.
The force on the mirror is
\begin{equation}
{\bf F} = \frac{d}{d\bar x_0}(2c{\bf G}_{total})
=\frac{d}{d\bar x_0}\int_V
2{\bbeta}\times{\bf \Pi} \, \delta (z)dv \, ,
\label{EQq10.08}
\end{equation}
which depends on the total momentum density, Eq.~(\ref{EQq10.04}).
If we were to assume ${\bf F}=2d{\bf G}_M/dt$, which is the relation
between momentum and force in an otherwise empty Minkowski spacetime,
then we would write 
\begin{equation}
{\bf F} = \frac{1}{c}\frac{d}{dt}\int_V
{2{\bf D}\times {\bf B}} \, \delta (z) dv \, .
\label{EQq10.09}
\end{equation}
Then one might use Eq.~(\ref{EQq10.09}) of interpret the results of
an experiment in such a way that the momentum density of the field
in the dielectric fluid is the Minkowski momentum density.
\par
The measured force on the mirror in the Jones--Richards
experiment \cite{BIExp}
is consistent with both Eqs.~(\ref{EQq10.08}) and Eqs.~(\ref{EQq10.09}),
depending on what theory you use to interpret the results.
Clearly an experiment that measures force, instead of directly
measuring the change in momentum in the dielectric, will not
conclusively distinguish the momentum density.
Specifically, the Jones--Richards experiment does not prove that the
Minkowski momentum density is the momentum density in the dielectric,
as has been argued, nor does it prove that the continuum momentum
density, Eq.~(\ref{EQq9.05}), is the momentum density in the dielectric.
\par
\section{Summary}
\par
It has been said that physics is an experimental science and that
physical theory must be constructed on the solid basis of
observations and measurements.
That is certainly true for serendipitous discoveries like x-rays and 
radioactivity; But Maxwell \cite{BIMaxwell} used inductive reasoning
to modify the Amp\`ere law and construct the laws of electrodynamics
two decades before Hertz \cite{BIHertz} demonstrated the existence
of electromagnetic waves.
Later, Einstein's theory of relativity was criticized for violating
the `well-established' principle of absolute
simultaneity \cite{BIEhrenfestQuote}.
The law of conservation of mass became the law of conservation of
mass--energy long before any measurements of relativistic mass
effects.
Mathematics is the language of physics and there are many other
examples (nonlinear optics, high-energy physics, negative refraction,
etc.) in which theoretical physics 
leads experiments by a substantial period of time.
\par
Axiomatic formal theory is a cornerstone of abstract mathematics.
The contradiction of valid theorems of Maxwellian continuum
electrodynamics proves, unambiguously, that Maxwellian continuum
electrodynamics is false.
Having proven Maxwellian continuum electrodynamics to be manifestly 
false, as it has been proven false by the Abraham--Minkowski momentum
contradiction for over a century, we established a reformulation of
theoretical continuum electrodynamics by deriving equations of motion
for the macroscopic fields from a generalized Lagrangian field theory.
For every simple linear medium there is a different set of equations
of motion based in a different continuous `material' spacetime with
coordinates that are renormalized by the linear permittivity and linear
permeability.
The Abraham--Minkowski controversy is trivially resolved because
the tensor total energy--momentum continuity theorem, the total
energy--momentum tensor, the total momentum, and the total energy are
fully electromagnetic, unique, and conserved for a closed (complete)
model system consisting of a simple linear dielectric block draped
with a gradient-index antireflection coating that is illuminated by
quasimonochromatic light.
\par
%\par
%\section*{Funding}
%\par
%\medskip
%\noindent The author is a US Government employee.
%\par
%\section*{Disclosures}
%\par
%\medskip
%\noindent The author declares no conflicts of interest.
%\par

\end{document}